\documentclass[twocolumn]{aastex63}

\newcommand{\swift}{{\it Swift}}
\received{\today}

\submitjournal{ApJ}

\shorttitle{Neutrino Counterparts}
\shortauthors{Franckowiak et al.}

\usepackage{xcolor}
\usepackage{xspace}

\newcommand{\PKS}{PKS 1502+106\xspace}
\newcommand{\TXS}{TXS 0506+056\xspace}
\newcommand{\HH}{1H 0323+342\xspace}
\newcommand{\MG}{MG3 J225517+2409\xspace}
\newcommand{\AT}{AT20G J175841-161703\xspace}

\usepackage{graphicx}
\usepackage{txfonts}

\begin{document}

\title{Patterns in the multi-wavelength behavior of candidate neutrino blazars}

\email{anna.franckowiak@desy.de, simone.garrappa@desy.de, vaidehi.s.paliya @gmail.com}

\author{A. Franckowiak}
\affiliation{Deutsches Elektronen-Synchrotron DESY, 15738 Zeuthen, Germany}

\author{S. Garrappa}
\affiliation{Deutsches Elektronen-Synchrotron DESY, 15738 Zeuthen, Germany}

\author{V. Paliya}
\affiliation{Deutsches Elektronen-Synchrotron DESY, 15738 Zeuthen, Germany}

\author{B. Shappee}
\affiliation{Institute for Astronomy, University of Hawai’i, Honolulu, HI 96822, USA}

\author{R. Stein}
\affiliation{Deutsches Elektronen-Synchrotron DESY, 15738 Zeuthen, Germany}

\author{N.~L. Strotjohann}
\affiliation{Department of Particle Physics and Astrophysics, Weizmann Institute of Science, Rehovot, Israel, 76100}

\author{M. Kowalski}
\affiliation{Deutsches Elektronen-Synchrotron DESY, 15738 Zeuthen, Germany}

\author{S. Buson}
\affiliation{University of W\"urzburg, 97074 W\"urzburg, Germany}

\author{S. Kiehlmann}
\affiliation{Institute of Astrophysics, Foundation for Research and Technology-Hellas, GR-71110 Heraklion,Greece}
\affiliation{Department of Physics, Univ. of Crete, GR-70013 Heraklion, Greece}

\author{W. Max-Moerbeck}
\affiliation{Departamento de Astronom\'{i}a, Universidad de Chile, Camino El Observatorio 1515, Las Condes, Santiago, Chile}

\author{R. Angioni}
\affiliation{ASI Space Science Data Center, Via del politecnico, snc. I-00133, Rome,Italy}
\affiliation{INFN Roma Tor Vergata, Via della Ricerca Scientifica, 1. I-00133, Rome, Italy}

\begin{abstract}

Motivated by the identification of the blazar \TXS as the first promising high-energy neutrino counterpart candidate, we search for additional neutrino blazars candidates among the \textit{Fermi}-LAT detected blazars.

We investigate the multi-wavelength behavior from radio to GeV gamma rays of blazars found to be in spatial coincidence with single high-energy neutrinos and lower-energy neutrino flare candidates. In addition, we compare the average gamma-ray emission of the potential neutrino-emitting sources to the entire sample of gamma-ray blazars. We find that neutrino-emitting blazar candidates are statistically compatible with both hypothesis of a linear correlation and of no correlation between neutrino and gamma-ray energy flux.

\end{abstract}


\section{Introduction}

After the detection of a diffuse flux of high-energy neutrinos~\citep{Aartsen:2013jdh} the most pressing challenge is to identify where these neutrinos are produced. Among the prime candidates are active galactic nuclei (AGN), especially those with a relativistic jet pointing towards us, so-called blazars \citep[e.g.][]{1991PhRvL..66.2697S,mannheim92,1993A&A...269...67M,szabo94,1995APh.....3..295M,mastichiadis96,protheroe99,atoyan01,dimitrakoudis12,Murase:2015ndr}. No significant clusters of neutrinos in either space or time have been identified by all-sky searches of IceCube data~\citep{Aartsen:2016oji,2015ApJ...807...46A,PhysRevLett.124.051103}. Searching for neutrinos from a predefined list of 110 sources revealed a $2.9\sigma$ excess at the position of the Seyfert II galaxy NGC 1068~\citep{PhysRevLett.124.051103}. Combining neutrino with multi-wavelength data is the key to probing neutrino emission from various source populations and to identifying potential electromagnetic counterparts. 

High-energy neutrinos are solely produced in the interaction of cosmic-ray nuclei with ambient matter or photon fields. In either case both charged and neutral pions are produced. The neutral pions decay into two gamma rays, the charged pions produce neutrinos in their decay chain. While gamma rays can also be produced in leptonic processes such as synchrotron emission, bremsstrahlung and inverse Compton scattering, neutrinos are exclusively produced in hadronic processes. They are therefore considered the smoking gun signature for the identification of cosmic-ray accelerators. Gamma rays produced alongside high-energy neutrinos can cascade down to lower energies through interactions within the source or during propagation. Increased neutrino activity might therefore be accompanied by increased electromagnetic emission that could appear in various wavelength bands.

The first likely extragalactic neutrino counterpart is the gamma-ray blazar \TXS, which was found to be in a flaring state in spatial and temporal coincidence with the arrival of the 290\,TeV neutrino event IC-190722A~\citep{MWScience} at $3\sigma$ significance. This finding motivated an archival search for lower $\mathcal{O}$(1-10\,TeV) neutrinos from the sky position of \TXS, which resulted in the detection of a 160\,day long neutrino flare in 2014/15 with $3.5\sigma$ significance~\citep{ICScience}. Surprisingly, this archival neutrino flare was not accompanied by increased activity in gamma-ray, optical or radio wavelengths~\citep{ICScience}. Note that no dedicated follow-up campaign was performed at the time of the neutrino flare and most of the available multi-wavelength data were collected by survey instruments. Hints for a hardening of the gamma-ray spectrum during the archival neutrino flare were identified by \citet{2018arXiv180704461P}, but were not found to be statistically significant ($\leq2\sigma$) by \citet{2019ApJ...880..103G}.

These two neutrino observations from the same source are difficult to reconcile through a single emission model: That the neutrino luminosity of the archival flare is more than four times larger than the gamma-ray luminosity~\citep{ICScience} suggests a hidden mechanism of neutrino production, e.g.\ through the attenuation of hadronic gamma rays due to cascades initiated by photons from the jet or the broad-line region \citep{2019ApJ...874L..29R,2019ApJ...881...46R}. We note that \citet{2019ApJ...874L..29R}, \citet{2019ApJ...881...46R} and \citet{2019arXiv191104010P} do not find a set of model parameters explaining the large neutrino flux from the archival neutrino flare without overshooting the electromagnetic observations. Furthermore, those hidden source scenarios~\citep{PhysRevLett.116.071101} are inconsistent with the association of IC-170922A with a strong gamma-ray flare from \TXS \citep[e.g.\ ][]{Gao:2018mnu,Ahnen:2018mvi,Cerruti:2018tmc}. There are however attempts to explain both observations in one single model by \citet{2019arXiv191011464Z} and \citet{2019PhRvD..99f3008L}, which come at the cost of assuming more complex geometries. The first one assumes a neutral beam scenario, while the second relies on hadronuclear interactions between protons in the jet and material in a dense gas cloud in the vicinity of the black hole. 

Accordingly, establishing and understanding either of the scenarios is of significant importance.

The detection of the archival neutrino flare from the direction of \TXS motivated a follow-up analysis~\citep{Erin} searching for similar neutrino flares from the position of all sources in the third catalog of AGN detected with the Large Area Telescope (LAT) on board the \textit{Fermi} Gamma-ray Space Telescope~\citep[3LAC,][]{Ackermann:2015yfk}. The most significant neutrino flare candidate was derived for each source, without accounting for electromagnetic observations of the sources. With neutrino data alone, no significant excess of flares was found above the expected atmospheric background. 

A similar situation occurred in the case of \TXS, where IC-170922A with a signalness of 56\% by itself was not significant and the archival neutrino flare in 2014/15 was not found significant in an all-sky neutrino only search. Only with the added information through multi-wavelength data was it possible to identify \TXS as the first promising candidate high-energy neutrino source.

The goal of this paper is to better understand blazars as possible source candidate for cosmic neutrinos and their emission mechanisms through the study of their electromagnetic activity. We search for coincidences of single well-reconstructed high-energy $\mathcal{O}$(100TeV) neutrino events with blazars detected by the \textit{Fermi} Large Area Telescope (LAT). Furthermore, we investigate the multi-wavelength behavior for the most significant sources reported by~\citet{Erin} and sources found spatially consistent with single high-energy neutrinos. 

We search for gamma-ray, X-ray, optical and radio activity correlated with the neutrino emission.

Finally, we study the ensemble of candidate sources by testing for generic properties expected for neutrino emitting source populations. 

We describe the sample of potential neutrino source candidates in Section~\ref{sec:SourceSample}. Section~\ref{sec:MWLC} describes the multi-wavelength data used to compile light curves and spectral energy distributions (SED) for this study and \ref{sec:methods} the statistical methods applied. We present our results in Section~\ref{sec:Results} and conclude in Section~\ref{sec:Conclusion}.

\section{Source sample}
\label{sec:SourceSample}
The neutrino sample used in this paper comes from the IceCube Neutrino Observatory, a cubic kilometer-scale Cherenkov detector located at the geographic South Pole. A complete description the IceCube detector is provided in ~\citep{2017JInst..12P3012A}.

\subsection{Neutrino flare candidates}
\citet{Erin} use a sample of well-reconstructed muon tracks from atmospheric and astrophysical neutrinos in the time period from April 26, 2012 to May 11, 2017. The sample covers the Northern sky at declinations above -5\,deg. The positions of 1023 sources from the 3LAC catalog are searched for a time-dependent neutrino signal in an unbinned maximum likelihood analysis. The eleven most significant neutrino flares are reported in \citet{Erin}. Two of the sources, B2 1126+37 and MG2 J112910+3702, are the two possible counterparts of 3FGL J1129.0+3705, which corresponds to 4FGL J1129.1+3703 in the fourth catalog of AGN detected by \textit{Fermi}-LAT \citep[4LAC,][]{Fermi-LAT:2019pir}. In 4LAC the source is associated only with CRATES J112916+370317, which has the same coordinates as MG2 J112910+3702. We therefore only keep CRATES J112916+370317 in our sample, i.e. we study the remaining ten sources reported by \citet{Erin}.

The temporal profile of each neutrino flare candidate is described by a Gaussian and for each flare the best fit central value $T_0$ and width $T_W$ of the Gaussian are reported. The latter is defined as twice the standard deviation of the Gaussian (see Table~\ref{tab:Sources}). \citet{Erin} report pre-trial p-values for the neutrino flare candidates ranging from $3.3\times10^{-3}$ to $3.5\times 10^{-5}$, but after trials correction none are significant. They perform a binomial test to assess the statistical significance of the ensemble yielding a p-value of 11\%, which increases to 24\% if \TXS is removed from the sample, compatible with expectations from background.

\subsection{Single high-energy neutrinos}
The IceCube realtime program selects high-energy ($\gtrsim$100\,TeV) starting and through-going muon track events~\citep{2017APh....92...30A}. A sample of realtime and archival events which would have qualified as a realtime alert, but was recorded before the realtime system was operational, was searched for blazar-neutrino coincidence~\citep{MWScience,2019ApJ...880..103G}. Both studies focused on well-reconstructed events with a $90\%$ containment radius of less than 5 square degrees. In addition to the coincidence of IC-170922A with \TXS, the $\sim100$\,TeV neutrino IC-141209A was identified in spatial coincidence with the BL Lac object GB6 J1040+0617. A detailed description of the multi-wavelength behavior of the two sources can be found in \citet{MWScience} and \citet{2019ApJ...880..103G} respectively. 
Here, we study the IceCube realtime alerts (see Tab.~\ref{tab:NeutrinoAlerts}) and archival neutrino events which would have passed the same selection criteria (see Tab.~\ref{tab:NeutrinoAlertsArchival}). 
Between April 2016 and May 2019, the IceCube collaboration operated two high-energy neutrino alert streams: the extremely high-energy (EHE) stream and the high-energy starting-track stream (HESE). In June 2019 the alert streams were unified to the GOLD and BRONZE streams defined by a purity of 50\% and 30\% respectively. Similarly to what was done in \citet{2019ApJ...880..103G}, we exclude sources with a 90\% angular uncertainty larger than 5 square degrees to remove events for which no significant association would be possible given their poor localization. Such events will typically be coincident with many blazars, resulting in a poor association probability. 

Between April 2016 and October 2019, a total of 35 alerts were issued, 16 survive our selection criteria (see Tab.~\ref{tab:NeutrinoAlerts}). Forty archival events have been identified between September 2010 and May 2016\footnote{Note that the EHE stream started only in July 2016, while the HESE stream was operational since April 2016.} and 28 pass our selection (see Tab.~\ref{tab:NeutrinoAlertsArchival}). Four additional coincidences are identified. The neutrino event IC-190730A was reported to be in spatial coincidence with the bright gamma-ray blazar \PKS~\citep{GCNPKS}.
Another coincidence was found with the high-energy starting track event IceCube-190221A and two 4FGL sources, 4FGL J1758.7-1621 and 4FGL J1750.4-1721. 
The first one is associated to a counterpart named \AT and classified as blazar of uncertain type (BCU), while the second one is un-associated. The neutrino best-fit position is located just 4 degrees from the Galactic plane, where the source density is high and also the large diffuse emission complicates the detection and association. Since our work focuses on AGN we only consider the BCU. 

In this work, we identify two additional new coincidences with archival neutrino events using 4LAC compared to the ones reported in \citet{2019ApJ...880..103G}, where 3LAC was used to search for coincidences. IC-150926A is spatially coincident with 4FGL J1258.7-0452 and IC-161103A with 4FGL J0244.7+1316. Both sources are included in 4LAC, but not in 3LAC. 

The ANTARES collaboration~\citep{ANTARES_ICRC} searched for an excess of neutrinos from the positions of 3LAC sources and reported a hot-spot of ANTARES neutrinos from the direction of \MG. \MG is also spatially coincident with the 340\,TeV neutrino IC-100608A with 65\% signalness \citep[event number three in][]{Aartsen:2016xlq} and in flaring state during the IceCube neutrino arrival time~\citep{ANTARES_ICRC}. We note that IC-100608A would not have passed our selection criteria outlined above due to its large 90\% angular uncertainty of roughly 30 square degrees (assuming an elliptical shape). Therefore, the source was excluded \textbf{from our} source sample test presented in Sec.~\ref{sec:KS}. However, the temporal coincidence with IC-100608A and spatial coincidence of the ANTARES hot-spot make this source interesting as a potential counterpart. 

\citet{2017MNRAS.466L..34K} reported a spatial coincidence between the HESE event IC-101112A and FSRQ PKS 0723-008. We do not consider this source here because it lies outside the reported 90\% uncertainty region\footnote{\url{https://icecube.wisc.edu/science/data/TXS0506_alerts}}, which is smaller than the originally published uncertainty radius in \citet{2014PhRvL.113j1101A}.

\begin{deluxetable*}{lllll m{5cm} c}
\tablecaption{Realtime neutrino alerts. Coordinates are reported in J2000 epoch with 90\% uncertainties. Alerts with a 90\% angular error larger than 5 square degrees are excluded from the analysis. For alerts printed in bold face a 4LAC source was identified located within the 90\% uncertainty region. The signalness is added for completion, but not used for further analysis.\label{tab:NeutrinoAlerts}}
\tablehead{
\colhead{IceCube Alert Name}  &
\colhead{Signalness} &
\colhead{Alert Type} & \colhead{RA[deg]} & \colhead{Dec[deg]} & \colhead{Coincident 4LAC source / Comments} & \colhead{GCN Circular}}  
\startdata
IC-191001A & 58.9\% & GOLD & $314.08^{+6.56}_{-2.26}$ & $12.94^{+1.50}_{-1.47}$ & large angular uncertainty & 25913 \\
IC-190922B & 50.5\% & GOLD & $5.76^{+ 1.19}_{-1.37}$ & $-1.57^{+ 0.93}_{-0.82}$ & -- & 25806\\
IC-190922A & 20.2\% &  GOLD & $167.43^{+3.40}_{-2.63}$ & $-22.39^{+2.88}_{-2.89}$ & large angular uncertainty & 25802\\
IC-190819A & 29.2\% & BRONZE & $148.80^{+2.07}_{-3.24}$ & $1.38^{+1.00}_{-0.75}$ & large angular uncertainty & 25402\\
\textbf{IC-190730A} & 67.2\% & GOLD & $225.79^{+1.28}_{-1.43}$ & $10.47^{+1.14}_{-0.89}$ & 4FGL J1504.4+1029 & 25225\\ 
IC-190712A & 30.3\% & BRONZE & $76.46^{+5.09}_{-6.83}$  & $13.06^{+4.48}_{-3.44}$ & large angular uncertainty & 25057\\
IC-190704A & 48.6\% & BRONZE & $161.85^{+2.16}_{-4.33}$ & $27.11^{+1.81}_{-1.83}$ & large angular uncertainty & 24981\\
IC-190629A & 33.9\% & BRONZE & $27.22$ & $84.33^{+4.95}_ {-3.13}$ & Dec value too close to pole for accurate error on RA & 24910\\
IC-190619A & 54.5\% & GOLD & $343.26^{+4.08}_{-2.63}$ & $10.73^{+1.51}_{-2.61}$ & large angular uncertainty & 24854\\
IC-190529A & 53\% & HESE & -- & -- & retracted & 24674  \\
IC-190504A & 63\% & HESE & $65.77$ & $-37.44$ & no detailed angular uncertainty provided~\citep{GCN24392} & 24392 \\
IC-190503A & 36.6\% & EHE & $120.28^{+0.57}_{-0.77}$ & $6.35^{+0.76}_{-0.70}$ & -- & 24378 \\
IC-190331A & 57\% & HESE & $337.68^{+0.23}_{-0.34}$ & $-20.70^{+0.30}_{-0.48}$ & -- & 24028\\
\textbf{IC-190221A} & 37\% & HESE & $268.81_{-1.8}^{+1.2}$ & $-17.04_{-0.5}^{+1.3}$ & 4FGL J1750.4-1721, 4FGL J1758.7-1621 & 23918 \\
IC-190205A & 84\% & HESE & -- & -- & retracted & 23876 \\
IC-190124A & 91\% & HESE & $307.40^{+0.8}_{-0.9}$ & $-32.18^{+0.7}_{-0.7}$ & -- & 23785 \\
IC-190104A & 35\% & HESE & $357.98^{+2.3}_{-2.1}$ & $-26.65^{+2.2}_{-2.5}$ &  -- & 23605 \\
IC-181031A & 87\% & HESE & -- & -- & retracted & 23398\\
IC-181023A & 28.0\% & EHE & $270.18^{+2.00}_{-1.70}$ & $-8.57_{-1.30}^{+1.25}$ & large angular uncertainty & 23375 \\
IC-181014A & 10\% & HESE & $225.15_{-2.85}^{+1.40}$ & $-34.80_{-1.85}^{+1.15}$ & large angular uncertainty & 23338\\
IC-180908A & 34.4\% & EHE & $144.58_{-1.45}^{+1.55}$ & $-2.13_{-1.2}^{+0.9}$ & -- & 23214 \\
IC-180423A & 34\% & HESE & -- & -- & retracted & 22669 \\
IC-171106A & 74.6\% & EHE & $340.00_{-0.50}^{+0.70}$ & $+7.40_{-0.25}^{+0.35}$ & -- & 22105\\
IC-171028A & 30\% & HESE & -- & -- & retracted & 22065 \\
IC-171015A & 51\% & HESE & $162.86_{-1.70}^{+2.60}$ & $-15.44_{-2.00}^{+1.60}$ & large angular uncertainty & 22016 \\
\textbf{IC-170922A} & 56.5\% & EHE & $77.43_{-0.80}^{+1.30}$ & $5.72_{-0.40}^{+0.70}$ & 4FGL J0509.4+0542 & 21916 \\ 
IC-170506A & 35\% & HESE & -- & -- & consistent with atmospheric muon background & 21075 \\
IC-170321A & 28.0\% & EHE & $98.30^{+1.2}_{-1.2}$ & $-15.02^{+1.2}_{-1.2}$ & -- & 20929 \\
IC-170312A & 78\% & HESE & $305.15^{+0.5}_{-0.5}$ & $-26.61^{+0.5}_{-0.5}$ & consistent with atmospheric muon background & 20857 \\
IC-161210A & 49.0\% & EHE & $46.58^{+1.10}_{-1.00}$ & $14.98^{+0.45}_{-0.40} $ & -- & 20247 \\
\textbf{IC-161103A} & 30\% & HESE & $40.83^{+1.10}_{-0.70}$ & $12.56^{+1.10}_{-0.65}$ & 4FGL J0244.7+1316 & 20119 \\ 
IC-160814A & 12\% & HESE & $200.3^{+2.43}_{-3.03}$ & $-32.4^{+1.39}_{-1.21}$ & large angular uncertainty & -- \\
IC-160806A & 28.0\% & EHE & $122.81^{+0.5}_{-0.5}$ &     $-0.81^{+0.5}_{-0.5}$ & -- & 19787 \\
IC-160731A & 84.9\% & EHE/HESE & $214.5^{+0.75}_{-0.75}$ & $-0.33^{+0.75}_{-0.75}$ & -- & -- \\
IC-160427A & 92\% & HESE & $240.57^{+0.6}_{-0.6}$ &     $9.34^{+0.6}_{-0.6}$ & -- & 19363 \\
\enddata
\tablecomments{Alerts taken from \url{https://gcn.gsfc.nasa.gov/amon_hese_events.html}, \url{https://gcn.gsfc.nasa.gov/amon_ehe_events.html} and \url{https://gcn.gsfc.nasa.gov/amon_icecube_gold_bronze_events.html}.}
\end{deluxetable*}

\begin{deluxetable*}{llllm{7cm}}
\tablecaption{Archival neutrino alerts. Coordinates are reported in J2000 epoch with 90\% uncertainties. Alerts with a 90\% angular uncertainty larger than 5 square degrees are excluded from the analysis. For alerts printed in bold face a 4LAC source was identified located within the 90\% uncertainty region. The signalness is added for completion, but not used for further analysis.\label{tab:NeutrinoAlertsArchival}}
\tablehead{
\colhead{IceCube Event Name}  &
\colhead{Alert Type} & \colhead{RA[deg]} & \colhead{Dec[deg]} & \colhead{Coincident 4LAC source / Comments}}  
\startdata
IC-160510A & EHE & $352.34^{+1.63}_{-1.31}$ &  $2.09^{+0.99}_{-0.85}$ & --\\
IC-160128A & EHE & $263.40^{+1.35}_{-1.18}$ & $-14.79^{+0.99}_{-1.02}$ & -- \\
IC-151207A & HESE & -- & -- & bad angular resolution would have been retracted\\
IC-151122A & EHE & $262.18^{+0.90}_{-1.21}$ & $-2.38^{+0.73}_{-0.43}$ & -- \\
\textbf{IC-150926A} & EHE & $194.50^{+0.76}_{-1.21}$ &  $-4.34^{+0.70}_{-0.95}$ & 4FGL J1258.7-0452\\
IC-150923A & EHE & $103.27^{+0.70}_{-1.36}$ & $3.88^{+0.59}_{-0.71}$ & --\\
IC-150911A & HESE & $240.20^{+1.29}_{-1.38}$ &  $-0.45^{+1.17}_{-1.23}$ & large angular uncertainty\\
IC-150831A & EHE & $54.85^{+0.94}_{-0.98}$ &  $33.96^{+1.07}_{-1.19}$ & -- \\
IC-150812A & EHE & $328.19^{+1.01}_{-1.03}$ &  $6.21^{+0.44}_{-0.49}$ & -- \\
IC-150428A & HESE & $80.77^{+1.12}_{-1.23}$ & $-20.75^{+0.45}_{-0.83}$ & --\\
\textbf{IC-141209A} & HESE & $160.05^{+0.84}_{-1.04}$ &   $6.57^{+0.64}_{-0.56}$ & 4FGL J1040.5+0617\\
IC-141109A & HESE & $55.63^{+0.79}_{-1.53}$ & $-16.50^{+0.81}_{-0.68}$ & no coincident sources\\
IC-140923A & EHE & $169.72^{+0.91}_{-0.86}$ &  $-1.34^{+0.73}_{-0.66}$ & -- \\
IC-140611A & EHE & $110.30^{+0.66}_{-0.45}$ &  $11.57^{+0.14}_{-0.24}$ & -- \\
IC-140420A & HESE & $238.98^{+1.81}_{-1.91}$ & $-37.73^{+1.47}_{-1.31}$ & large angular uncertainty \\
IC-140203A & EHE & $349.54^{+2.21}_{-1.97}$ & $-13.71^{+1.23}_{-1.38}$ & large angular uncertainty\\
IC-140122A & HESE & $219.64^{+5.16}_{-4.16}$ & $-86.16^{+0.55}_{-0.60}$ & large angular uncertainty\\
IC-140109A & EHE & $292.85^{+0.87}_{-0.94}$ &  $33.06^{+0.50}_{-0.46}$ & -- \\
IC-140108A & EHE & $344.53^{+0.67}_{-0.48}$ & $1.57^{+0.35}_{-0.32}$ & -- \\
IC-131204A & EHE & $289.16^{+1.08}_{-0.94}$ & $-14.25^{+0.91}_{-0.81}$ & -- \\
IC-131202A & HESE & $206.63^{+2.04}_{-1.56}$ & $-22.02^ {+1.69}_{-1.04}$ & large angular uncertainty \\
IC-131023A & EHE & $301.82^{+1.10}_{-0.93}$ &  $11.49^{+1.19}_{-1.09}$ & -- \\
IC-130907A & EHE & $129.81^{+0.48}_{-0.28}$ & $-10.36^{+0.36}_{-0.31}$ & --\\
IC-130627A & HESE & $93.43^{+0.80}_{-0.85}$ &  $14.02^{+0.72}_{-0.75}$ & no coincident sources\\
IC-130408A & HESE & $167.17^{+2.87}_{-1.90}$ &  $20.67 ^{+1.15}_{-0.89}$ & large angular uncertainty\\
IC-121011A & EHE & $205.22^{+0.59}_{-0.65}$ &  $-2.39^{+0.51}_{-0.57}$ & -- \\
IC-120922A & EHE & $70.75^{+1.56}_{-1.63}$ &  $19.79^{+1.37}_{-0.68}$ & large angular uncertainty\\
IC-120523A & EHE & $171.03^{+0.81}_{-0.90}$ & $26.36^{+0.49}_{-0.30}$ & -- \\
IC-120501A & HESE & -- & -- & bad angular resolution would have been retracted\\
IC-120301A & EHE & $238.01^{+0.60}_{-0.59}$ &  $18.60^{+0.46}_{-0.39}$ & -- \\
IC-111228A & HESE & -- & -- & bad angular resolution would have been retracted\\
IC-110930A & EHE & $266.48^{+2.09}_{-1.55}$  & $-4.41^{+0.59}_{-0.86}$ & -- \\
IC-110714A & HESE & $67.86^{+0.51}_{-0.72}$ & $40.32^ {+0.73}_{-0.25}$& -- \\
IC-110304A & EHE & $116.37^{+0.73}_{-0.73}$ & $-10.72^{+0.57}_{-0.65}$ & -- \\
IC-110216A & HESE & -- & -- & bad angular resolution would have been retracted \\
IC-110128A & EHE & $307.53^{+0.82}_{-0.81}$ &  $1.19^{+0.35}_{-0.32}$ & -- \\
IC-101112A & HESE & $110.56^{+0.80}_{-0.37}$ & $-0.37^{+0.48}_{-0.65}$ & -- \\
IC-101028A & EHE & $88.68^{+0.54}_{-0.55}$ &  $0.46^{+0.33}_{-0.27}$ & -- \\
IC-101009A & EHE & $331.09^{+0.56}_{-0.72}$ & $11.10^{+0.48}_{-0.58}$ & --\\
IC-100912A & HESE & -- & -- & bad angular resolution would have been retracted \\
\enddata
\tablecomments{Archival events taken from \url{https://icecube.wisc.edu/science/data/TXS0506_alerts}}
\end{deluxetable*}

\section{Multi-wavelength Data}
\label{sec:MWLC}

In the following we motivate why different wavelengths may provide relevant information connected to high-energy neutrino emission.

High-energy neutrinos are produced together with high-energy photons of similar energy in hadronic processes. TeV to PeV photons are quickly absorbed within the source or in interactions with the extragalactic background light through photon-photon annihilation and cascade down to lower energies. Hence, GeV gamma rays detected by \textit{Fermi}-LAT provide the all-sky dataset closest in energy to the neutrinos of interest. However, if the source environment is optically thick to GeV gamma rays due to high densities of photons in the keV range, then those gamma rays will cascade down to even lower energies, which then become an important tracer of the source activity as well. Furthermore, TeV instruments relying on the imaging atmospheric Cherenkov technique have a limited field of view and therefore even with archival data do not provide an all-sky coverage. All-sky TeV instruments such as HAWC~\citep{Abeysekara:2013tza} have limited sensitivity to extragalactic sources due to EBL absorption.

X-rays might be a good tracer for hadronic interactions in sources where the GeV emission is dominated by leptonic processes~\citep{Gao:2018mnu,Keivani:2018rnh}.

Increased radio emission was found from \TXS at the arrival time of IC-170922A and PKS B1424-418 in coincidence with the arrival time of a PeV neutrino~\citep{2016NatPh..12..807K}. We note that the chance coincidence of the neutrino association with the latter was relatively large ($5\%$). \citet{2019A&A...630A.103B} use radio data to suggest a possible collision of two jets in \TXS. However, \citet{2019arXiv191201743R} exclude the presence of a secondary jet core with higher resolution radio data, but find signs of a spine-sheath structure of the jet, which could be relevant for neutrino production see also \citep[see also][]{2005A&A...432..401G,2014ApJ...793L..18T,Ahnen:2018mvi}.

X-ray and gamma-ray polarization data could be used to pinpoint the leptonic and/or hadronic blazar radiation mechanisms in the high-energy bands, and to infer the magnetic field strength in the emission region~\citep{Zhang:2019dob}, but are not available for sources in our sample.

Finally, archival optical data are available for all sources of our sample. In combination with gamma-ray data, optical data can be useful to identify high-energy flares without low-energy counterparts, which could be due to hadronic interaction~\citep{2004ApJ...601..151K}.

\subsection{\textit{Fermi}-LAT data}

The \textit{Fermi}-LAT is a pair-conversion telescope sensitive to gamma rays with energies from $20\,$MeV to greater than $300\,$GeV~\citep{2009ApJ...697.1071A}. It has a field of view $>$ 2sr and scans the entire sky every three hours during standard operations.
We use almost 11 years of Pass\,8 data collected between 2008 August 4 and 2019 May 30 (MJD 54682-58633) with an exception for the source \PKS for which we use data up to 2019 July 31 (MJD 58695) in order to include the arrival time of IC-190730A. We select photons from the event class developed for point source analyses\footnote{\url{http://fermi.gsfc.nasa.gov/ssc/data/analysis/documentation/Pass8_usage.html}} in the energy range from 100\,MeV to 800\,GeV binned into ten logarithmically-spaced energy intervals per decade.
We select a region of interest (ROI) of $15\degr \times 15\degr$ centered on the gamma-ray source position, binned in $0\fdg1$ size pixels. The binning is applied in celestial coordinates using a Hammer-Aitoff projection. We perform a maximum likelihood analysis using the standard \textit{Fermi}-LAT ScienceTools package version v11r04p00 available from the \textit{Fermi} Science Support Center\footnote{\url{http://fermi.gsfc.nasa.gov/ssc/data/analysis/}} (FSSC) and the \textsf{P8R3\_SOURCE\_V2} instrument response functions, together with the fermipy package v0.17.4 \citep{Wood:2017yyb}.

We use standard data-quality cuts to select events observed when the detector was in a normal operation mode.  
In order to obtain a sample of events for each analysis with a reduced contamination from gamma rays produced in the Earth's upper atmosphere, we apply an additional instrument zenith angle cut of $\theta<90\degr$. We also remove time periods coinciding with bright solar flares and gamma-ray bursts detected by the LAT.  
The input model for the ROI includes all known gamma-ray sources from the 4FGL catalog in a region of $20\degr \times 20\degr$, slightly larger than the ROI, and the isotropic and Galactic diffuse gamma-ray emission models provided by the standard templates \textsf{iso\_P8R3\_SOURCE\_V2\_v01.txt} and \textsf{gll\_iem\_v07.fits}\footnote{\url{https://fermi.gsfc.nasa.gov/ssc/data/access/lat/BackgroundModels.html}}. The effect of energy dispersion is included in the fits performed with the \textit{Fermi}-LAT ScienceTools for all point sources and the Galactic diffuse gamma-ray emission model. 
We use an iterative source-finding algorithm to scan the ROI and include in the model sources that are significantly ($\geq 5\,\sigma$) detected over the full dataset time range, but not over the 8-year data that produced the 4FGL catalog. New putative point sources are modeled with a single power-law spectrum, with the index fixed to 2 and the normalisation free to vary in the fit. The search procedure is iterated until no further significant excess is found. The new point
sources significantly detected in the longer-integration
time data set are accounted for by the final ROI model.

The definition of test statistics (TS) from \citet{mattox1996} is used to measure the detection level of each source. The minimum separation allowed between two independent point source detections is set to $0\fdg3$.  
We compute the light curve for each source using the adaptive binning algorithm from \citep{2012A&A...544A...6L} with the prescriptions outlined in \citep{2019ApJ...880..103G}, in order to better resolve flaring activities of the target sources. Statistically-significant variations in the light curve's behavior are detected in this work with the Bayesian Blocks algorithm \citep{2013ApJ...764..167S} for which we use its Astropy implementation\footnote{\url{http://docs.astropy.org/en/stable/api/astropy.stats.bayesian_blocks.html}}. We adopt a prior that makes the algorithm sensitive to variations that are significant at 95\% confidence level.

All reported gamma-ray fluxes are in the analysis energy range from 100\,MeV to 800\,GeV.

\subsection{\textit{Neil Gehrels Swift observatory} data}
While no sensitive all-sky X-ray monitor exists, we can take advantage of pointed observations of the \textit{Swift} X-Ray Telescope (XRT) collected in target of opportunity and monitoring operations. We first reprocessed the \swift-XRT data to calibrate and clean the event files using the task {\tt xrtpipeline}. 

The pipeline {\tt xrtgrblc} was adopted to extract the source and background spectra and ancillary response files used for the light curve generation. This tool automatically adjusts the source and background region sizes based on the source count rate\footnote{https://heasarc.gsfc.nasa.gov/lheasoft/ftools/headas/xrtgrblc.html}. Due to low photon statistics of the individual observation ids, we fit a simple absorbed power law model in XSPEC \citep[][]{1996ASPC..101...17A}, while taking the Galactic neutral hydrogen column density along the line of sight from \citet[][]{2005AnA...440..775K}.

For the broadband SEDs, the event files were combined with {\tt xselect}. Exposure maps and ancillary response files were extracted with the tasks {\tt ximage} and {\tt xrtmkarf}. The source region was chosen as a circle of 47$^{\prime\prime}$ radius centered at the target, whereas the background region has an annular shape with inner and out radii of 70$^{\prime\prime}$ and 150$^{\prime\prime}$, respectively, centered at the source of interest. We tested both an unbroken and broken power law taking into account the Galactic neutral hydrogen column density along the line of sight \citep[][]{2005AnA...440..775K} and report the spectral parameters for the model which represents the data better. Depending on the source brightness, the source spectra are re-binned to have at least 20 or 1 counts per bin. The spectral analysis is performed in XSPEC.

Snapshot observations from the UltraViolet and Optical Telescope (UVOT) on-board the {\it Swift} satellite during each pointing to the target source are first combined using the tool {\tt uvotimsum}. To derive the source instrumental magnitude using {\tt uvotsource}, we adopt a circular source region of 5$^{\prime\prime}$ radius centered at the object position and a nearby source-free region of 30$^{\prime\prime}$ radius is considered to derive the background contamination. The computed magnitudes are converted to energy flux units using the zero points and calibrations of \citet[][]{2011AIPC.1358..373B} corrected for the Galactic reddening following \citet[][]{2011ApJ...737..103S}.

\subsection{\textit{ASAS-SN} and CSS optical data}

Optical data in the V-band and g-band 
from the All-Sky Automated Survey for Supernovae \citep[ASAS-SN,][]{2014ApJ...788...48S,2017PASP..129j4502K} are processed by the fully automatic ASAS-SN pipeline using the ISIS image subtraction package \citep{alard98, alard00}. We then perform aperture photometry on the subtracted science image using the IRAF {\tt apphot} package, adding back in the flux from the reference image. The photometry is calibrated using the AAVSO Photometric All-Sky Survey (APASS, \citealp{henden15}). 

Additional V-band data from the Catalina Sky Survey ~\citep[CSS,][]{2009ApJ...696..870D} are available from the public database and are based on aperture photometry. To mitigate color-dependent differences between the UVOT, CSS and ASAS-SN V-band filters, we add an offset to the ASAS-SN and UVOT data to match the CSS data in regions with overlapping exposure. A similar offset was applied to the ASAS-SN g-band observations to line them up with the V-band data. The applied shift is a constant in flux space, and is indicated in the legend of the corresponding light curve figures. Since our study only relies on the shape of the light curve rather than the absolute optical flux level, and given that none of the neutrino flares occurred in the transition region between ASAS-SN and CSS data, this shift is not critical for our results.

\subsection{Radio Data}
Owens Valley Radio Observatory (OVRO) 15GHz radio monitoring data~\citep{OVRO} are available for nine sources of the sample \citep[one of them is \TXS which was already presented in][]{MWScience}.

\subsection{Other Data}
We collect archival spectral observations with the Space Science Data Center SED builder tool\footnote{https://tools.ssdc.asi.it/} to supplement the data analyzed in this work. This allows us to cover the broadband SED of the target objects as well as possible, admittedly using non-simultaneous data sets. However, considering that these observations represent an `average' activity of the sources, we can use them to compare the existing data acquired contemporaneously to the reported neutrino events.

\begin{deluxetable*}{l l l l l l l l l}
\tablecaption{Neutrino source candidates. The first seven sources were found in coincidence with single high-energy neutrinos, while the remaining sources were found coincident with neutrino flares by~\citet{Erin}.\label{tab:Sources}}
\tablehead{
\colhead{Source Name}  & \colhead{4FGL Name} & \colhead{Class} & \colhead{redshift} & \colhead{$T_0$ [MJD]} & \colhead{$T_{w}$ [days]} & \colhead{$p_\gamma$} & \colhead{$T_{\gamma,\nu}$[MJD]} & \colhead{$L_\gamma$ [erg/s]}
}  
\startdata
\multicolumn{9}{c}{Single high-energy neutrinos}\\
\hline
   \MG  & J2255.2+2411  & BL Lac  
   & 1.37\tablenotemark{a} & 55355.49 & -- & 0.04 & [55346.73, 55403.54] & $1.3 \times 10^{47}$\\
   GB6 J1040+0617   & J1040.5+0617 & BL Lac 
   & 0.73\tablenotemark{b} & 57000.14311 & -- & 0.17 & [56997.67, 57055.08] & $4.6 \times 10^{46}$\\
   1RXS J125847.7-044746 & J1258.7-0452 & BL Lac & 0.586\tablenotemark{c} & 57291.90119 & -- & -- & -- & $2.9 \times 10^{45}$\\
   GB6 J0244+1320 & J0244.7+1316  & BCU\tablenotemark{d} & -- & 57695.38 & -- & -- & -- & -- \\
   \TXS & J0509.4+0542  & BL Lac\tablenotemark{e}                
   & 0.336\tablenotemark{f} & 58018.87 &  -- & 0.009 & [58016.57, 58019.94] & $2.2 \times 10^{46}$ \\  
   \AT & J1758.7-1621 & BCU 
   & - & 58535.35 & -- & 0.39 & [58304.43, 58633.01] & -- \\
   \PKS  & J1504.4+1029 & FSRQ    
   & 1.839 & 58694.8685 & -- & 0.75 & [58603.54, 58695.14] & $4.7\times 10^{48}$\\
\hline
\multicolumn{9}{c}{Neutrino flare candidates}\\
\hline
   4C +20.25 & J1125.9+2005 & FSRQ                  
   & 0.133 &56464.1    & $5.2$ & 0.64 & [56369.45, 57248.31] & $1.6 \times 10^{44}$\\
   CRATES J112916+370317 & J1129.1+3703 & BL Lac                
   & 0.445 & 56501.385 & $6.0\times10^{-2}$ & 0.45 & [56404.68, 57066.59] & $2.9\times 10^{46}$\\
   MG2 J112758+3620 & J1127.8+3618& FSRQ 
   & 0.884 & 56501.385 & $6.0\times10^{-2}$ & 0.24 & [56482.90, 56555.93] & $5.5\times 10^{46}$\\
   \TXS & J0509.4+0542  & BL Lac\tablenotemark{e}                 
   & 0.336 & 57000 &  $120$ & 0.92 & [56965.28, 57089.28] & $2.2 \times 10^{46}$\\  
   \HH  & J0324.8+3412 & NLSY1\tablenotemark{g} 
   & 0.061 & 57326.2938 & $1.7\times10^{-3}$ & 0.08 & [57326.10, 57333.17] & $2.0\times 10^{44}$\\
   RBS 1467 & J1508.8+2708	& BL Lac                
   & 0.27 &57440      & $170$ & 0.53 & [56474.88, 58633.01] & $6.3 \times 10^{44}$\\
   S4 1716+68 & J1716.1+6836 & FSRQ                  
   & 0.777 &57469.17919 & $5.4\times10^{-5}$ & 0.48 & [57378.18, 57510.76] & $2.1\times 10^{46}$\\
   M 87 & J1230.8+1223	& radio galaxy          
   & 0.00428 & 57730.0307& $2.7\times10^{-3}$ & 0.55 & [57724.77, 57847.51] & $6.9\times 10^{41}$\\
   GB6 J0929+5013 & J0929.3+5014  & BL Lac                
   & 0.37\tablenotemark{h} & 57758.0   & $1.2$ & 0.44 & [57647.78, 57759.66] & $5.6 \times 10^{45}$\\
   1ES 0927+500 &J0930.5+4951	& BL Lac                
   & 0.187 &57758.0    & $1.2$ & 0.49 & [57031.36, 58633.01] & $2.2 \times 10^{44}$\\
\enddata
\tablecomments{The source classes and redshifts (if not noted otherwise) are reported in the 4LAC catalog~\citep{Fermi-LAT:2019pir}. $T_0$ is the central time of the reported neutrino flare and $T_{w}$ is twice the standard deviation of the Gaussian flare. $p_\gamma$ is the probability that the neutrino flare center is coincidence with a gamma-flare of the found or larger flux. $T_{\gamma,\nu}$ is the time window used to calculate the gamma-ray flux in which the neutrino arrived. The gamma-ray luminosity is the 8-year average calculated from the 4LAC values. Luminosity is only calculated when a redshift measurement is available. 1RXS J125847.7-044746 and GB6 J0244+1320 are too dim in gamma rays to study the variability.}
\tablenotetext{a}{Redshift from 4LAC, which is taken from SDSS, where it is flagged as ``chi-squared of best fit is too close to that of second best ($< 0.01$ in reduced chi-squared)''. \citet{PaianoATel} find that the redshift is $>0.8633$.}
\tablenotetext{b}{Redshift from \citet{GB_redshift_old}.}
\tablenotetext{c}{Redshift from \citet{2000ApJS..129..547B}.}
\tablenotetext{d}{Blazar of uncertain type.}
\tablenotetext{e}{Note that \TXS was re-classified by \citet{Padovani:2019xcv} as ``masquerading BL Lac'', i.e., intrinsically a flat-spectrum radio quasar with hidden broad lines and a standard accretion disk.}
\tablenotetext{f}{Redshift from~\citet{Paiano:2018qeq}.}
\tablenotetext{g}{Narrow line Seyfert 1.}
\tablenotetext{h}{Redshift from \citet{2009yCat.2294....0A}.}
\end{deluxetable*}

\section{Methods}
\label{sec:methods}

\subsection{Quantifying the gamma-ray activity during the neutrino arrival time}
For each individual source we calculate the chance probability 
\begin{equation}
    p_\gamma (F_{\gamma,\nu}) = \frac{\int^{\infty}_{-\infty} \left( \sum\limits_{i}t_i \int^{\infty}_{F_{x}} \mathcal{N}(x,F_i,\sigma_i) dx \right) \mathcal{N}(F_x,F_{\gamma,\nu},\sigma_{\gamma,\nu}) dF_x}{\sum\limits_{i} t_i}
\end{equation}
to find the neutrino in a period of gamma-ray activity, larger than the gamma-ray energy flux, $F_{\gamma,\nu}$ in the time bin, $t$, overlapping with the neutrino arrival. Here $\mathcal{N}$ is a Gaussian function with mean $F_{\gamma,\nu}$ and standard deviation $\sigma_{\gamma,\nu}$ evaluated at $F_x$, i.e. we assume that the flux uncertainty is normally distributed. The index, $i$, runs over all time bins of the source of interest. $p_\gamma$ for all sources is reported it in Tab.~\ref{tab:Sources}. Low values of $p_\gamma$ indicate that the source was in a high gamma-ray flux state during the neutrino arrival time compared to the other time bins in the 11-year light curve, while high values indicate that the source did not show an excess in gamma rays in temporal coincidence with the neutrino emission. We use the adaptive bins that were used to compile the gamma-ray light curves. Due to non-continuous exposure and gaps in the data we do not perform a similar analysis for optical and X-ray data. We note that the optical data show in general a similar temporal behavior to the gamma-ray data, as was found in previous studies \citep[see e.g.][]{Cohen:2014bja}.

\subsection{Comparison of neutrino blazar candidates to the gamma-ray blazar sample}
\label{sec:KS}

\begin{figure*}
    \centering
           \includegraphics[width=14cm]{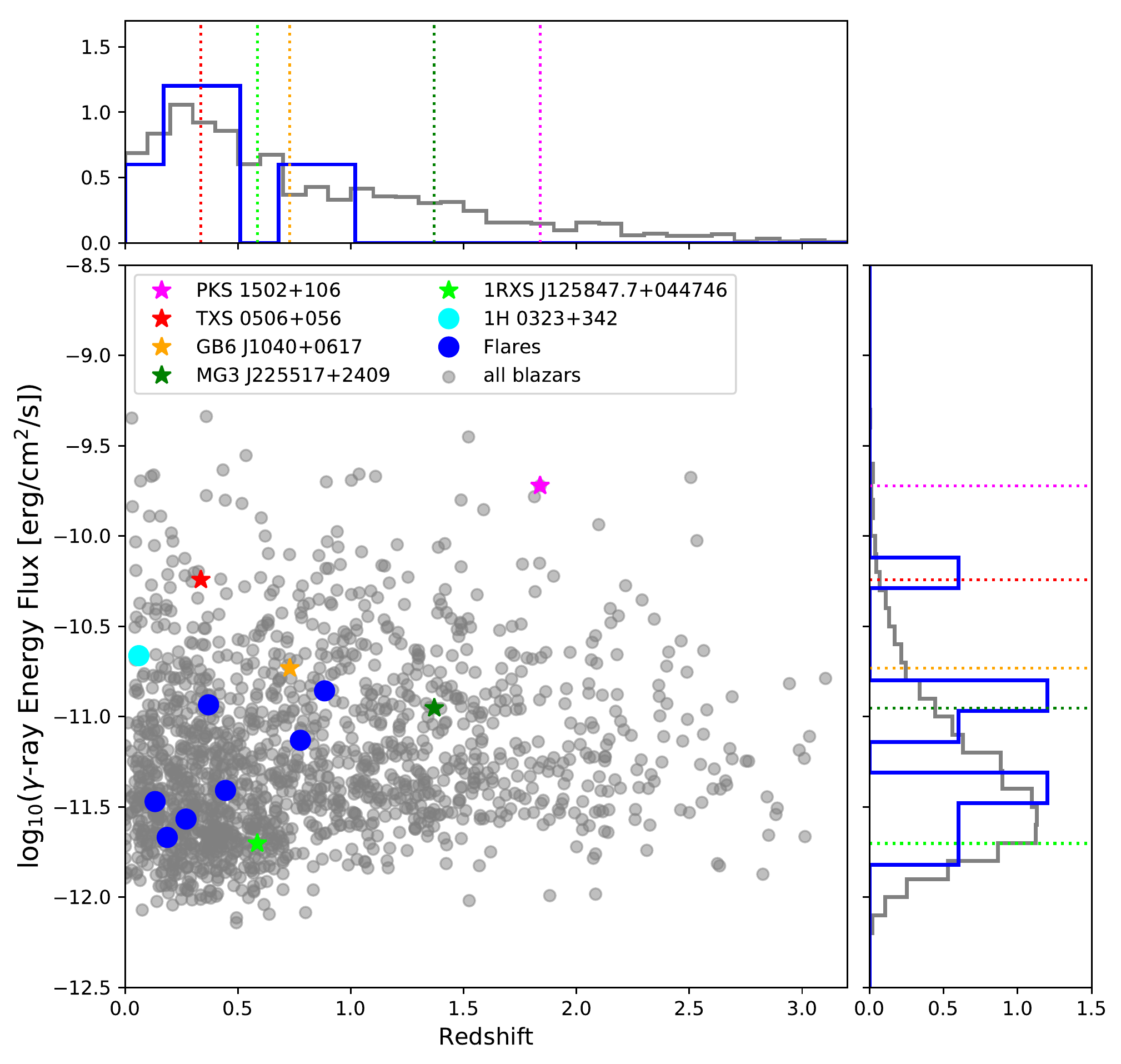}
     \caption{Comparison of candidate neutrino blazars with all blazars in the 4LAC AGN sample (shown in grey). The gamma-ray energy flux is shown as a function of redshift. Sources identified in the neutrino flare search are displayed by blue circles. \HH is high-lighted in cyan. Sources associated with single high-energy neutrinos are marked by colored stars. The side panels show projections of the distributions. The dashed lines in the projection panels are the values of individual blazars associated with single high-energy neutrinos and the blue distribution shows the histogram of the neutrino flare candidate sources.
              }
         \label{fig:catalogAll}
\end{figure*}

In addition to studying the multi-wavelength behavior of individual sources, we study the average gamma-ray properties of the sources identified as potential neutrino emitters and compare them to the entire gamma-ray blazar population. Figure \ref{fig:catalogAll} shows the time-integrated gamma-ray energy flux in the energy range from 100\,MeV to 100\,GeV as a function of redshift for all blazars in 4LAC (including blazars of uncertain type). All values are taken from 4LAC. We have added the redshift of four sources (see Tab.~\ref{tab:Sources}).  
We apply a Kolmogorov-Smirnov (KS) test to determine how compatible the gamma-ray energy flux distribution of the candidate neutrino blazars is with the expected distribution of gamma-ray blazars under a given hypothesis. 
To verify that the KS test p-value is not biased, we performed a sanity check with randomized data. We generate a background KS p-value distribution by randomly selecting $N$ blazars from the entire blazar sample and calculating the KS p-value for those. $N$ is the number of identified neutrino blazar candidates. A calibrated p-value for the measurement is then calculated as the ratio of background p-values smaller than the measured KS p-value. We note that no significant bias was found and the calibrated p-value is similar to the one obtained directly from the KS test method.

We compare the observed gamma-ray energy flux of candidate neutrino blazars to the expectation for three separate scenarios. First, we test the uncorrelated case, in which all neutrino blazar coincidences occur by chance. In that case we expect the gamma-ray flux of the candidate neutrino blazars to follow the distribution of the gamma-ray blazar population as a whole. Second, we test the hypothesis of a linear correlation of the neutrino flux with the gamma-ray energy flux of blazars. In that case we expect the candidate neutrino blazars to have preferentially higher gamma-ray energy flux. In the third case, we test whether the neutrino flux is proportional to the square of the gamma-ray energy flux as has been suggested in \citet{Oikonomou:2019pmg}. Here, we expect the candidate neutrino blazar distribution to be skewed towards even higher gamma-ray energy fluxes.

For internal consistency, the single high-energy neutrino blazar candidates are compared to the 4LAC blazar population, while the neutrino flare blazar candidates are compared to 3LAC, because only 3LAC source positions were searched for neutrino flares in \citet{Erin}.
A large KS p-value implies that the data is well described by a given hypothesis, while a small one indicates that that hypothesis is disfavored.

The same KS test is also applied to the candidate neutrino-flare sources. As pointed out in \citet{Erin}, one pair and one triplet of sources are correlated. After associating 4FGL J1129.1+3703 with CRATES J112916+370317, the triplet becomes a pair, because the second possible counterpart can be discarded. The position of CRATES J112916+370317 is correlated with MG2 J112758+3620, which is associated to the gamma-ray source 4FGL J1127.8+3618. 
The second correlated source positions are GB6 J0929+5013 and 1ES 0927+500 (associated to the gamma-ray sources 4FGL J0929.3+5014 and 4FGL J0930.5+4951 respectively). We recalculate the KS test using only one of the correlated source positions and quote a range of p-values, which brackets the outcome of removing a different set of sources from the test. 

\MG did not fulfill our angular uncertainty criteria and was therefore excluded from the KS test.

The results of the KS test are presented in Tab.~\ref{table:KS} split into BL Lacs, FSRQs and all blazars (including also blazars of uncertain type) combined.

\begin{deluxetable*}{c c c c c c c}
\tablecaption{KS test p-values. The range of p-values for the neutrino flare case comes from removing different combinations of the correlated source positions. Different columns represent the uncorrelated and linearly-correlated hypothesis, values in parentheses represent the quadratically-correlated case.
Note that neutrino flare candidate blazars are compared to the 3LAC population and single high-energy neutrino candidate blazars with the 4LAC population.\label{table:KS}}
\tablehead{
 & \multicolumn{2}{c}{BL Lacs}  & \multicolumn{2}{c}{FSRQs} & \multicolumn{2}{c}{All Blazars} \\    
 & uncorrelated & correlated & uncorrelated & correlated & uncorrelated & correlated
}  
\startdata
   Single Neutrinos & 0.32 & 0.45 (0.0013) & 0.10 & 0.36 (0.28) & 0.126 & 0.64 (0.00032) \\
   Neutrino Flares&  0.37-0.98 & 0.027-0.533 & 0.01-0.36 & 0.0075-0.023 & 0.39-0.98 & 0.0039-0.021\\
\enddata
\end{deluxetable*}

\section{Results}
\label{sec:Results}

\subsection{Individual sources} 

The collected multi-wavelength light curves are presented in multi-panel Fig.~\ref{fig:LC_MG3J225517+2409} to Fig.~\ref{fig:1ES0927+500} in the appendix. We do not show the light curves of \TXS and GB6 J1040+0617, because they were already discussed in detail in~\citet{2019ApJ...880..103G}.
 
All sources are detected in GeV gamma rays, which is expected since they are selected from the 3LAC or 4FGL catalog. However, some of them are too faint to resolve temporal structure. We present both the flux variation and the spectral index variation assuming a power-law spectrum for the source in each bin. 

Most sources have a good coverage in optical during the neutrino arrival times. Radio data from the OVRO monitoring program is available for nine out of 14 sources. 
X-ray data is sparse and only available for eight sources. Only \HH has a good coverage in X-rays during the neutrino flare. 

In the following we discuss the three most interesting sources. We discuss the brightest source in gamma rays, \PKS, and the two sources, \HH and \MG, which show gamma-ray flares during the neutrino arrival time, reflected by small $p_\gamma$ of $8\%$ and $4\%$ respectively, while the other sources showed p-values ranging from 17\% to 92\%. However, given that we have performed this calculation for fifteen sources, these findings are well compatible with the background expectations.

\subsubsection{\HH}
 
\begin{figure*}
    \centering
           \includegraphics[width=13cm]{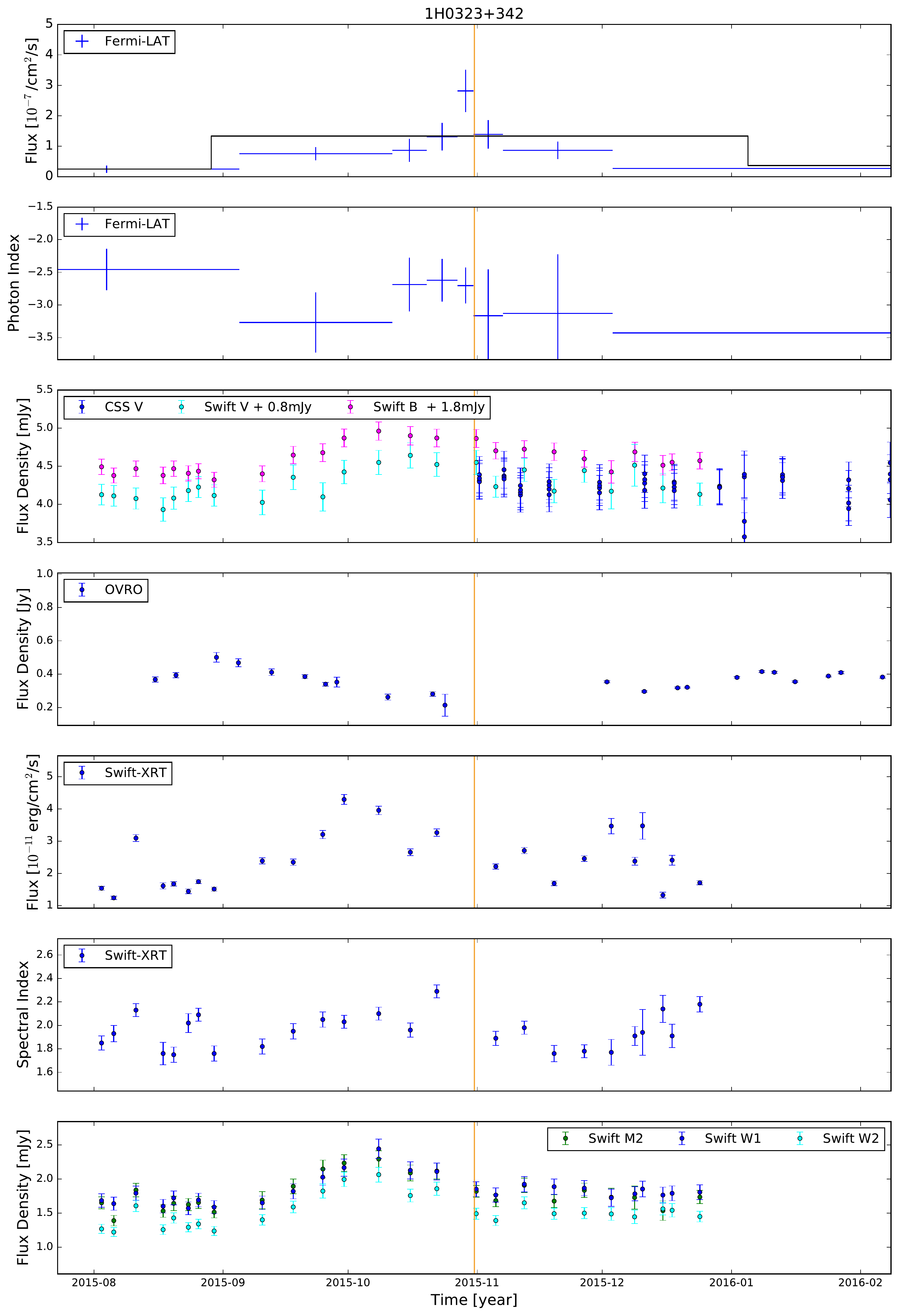}
     \caption{Multi-wavelength light curve of 1H 0323+342. The duration of the neutrino flare is short ($T_w$ = 147\,s) and its arrival time is shown as an orange line. An excess in gamma rays is found coincident with the neutrino arrival time and an excess in X-ray emission is visible roughly one month before the neutrino arrival time. The \textit{Fermi}-LAT gamma-ray light curve covers the energy range from 100\,MeV to 800\,GeV, the \textit{Swift} X-ray light curve from 0.3 to 10\,keV and the OVRO radio data is at 15\,GHz.
              }
         \label{fig:LC_1H0323+342_zoom}
\end{figure*}

\begin{figure*}
    \centering
           \includegraphics[width=8.5cm]{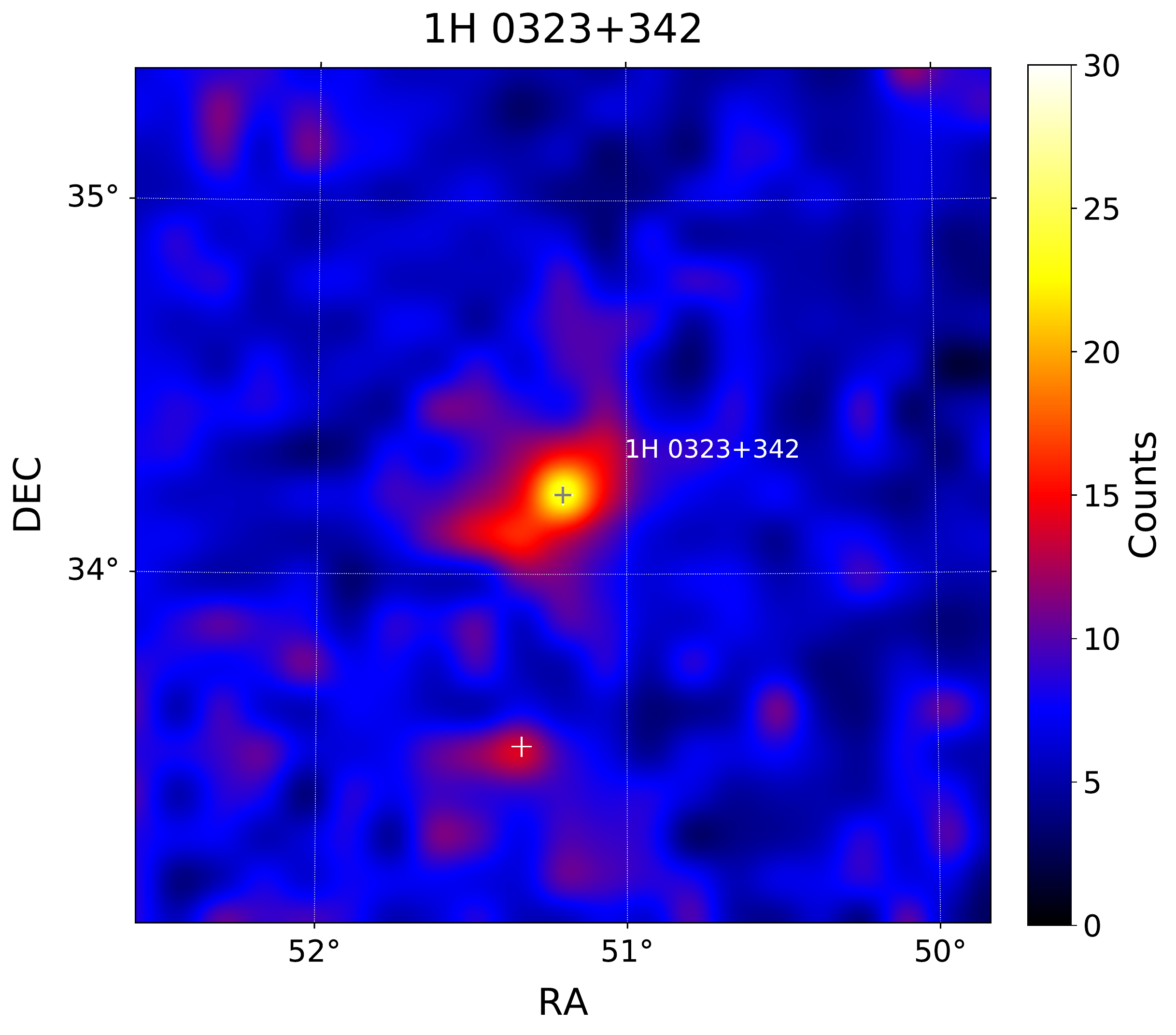}
           \includegraphics[width=8.5cm]{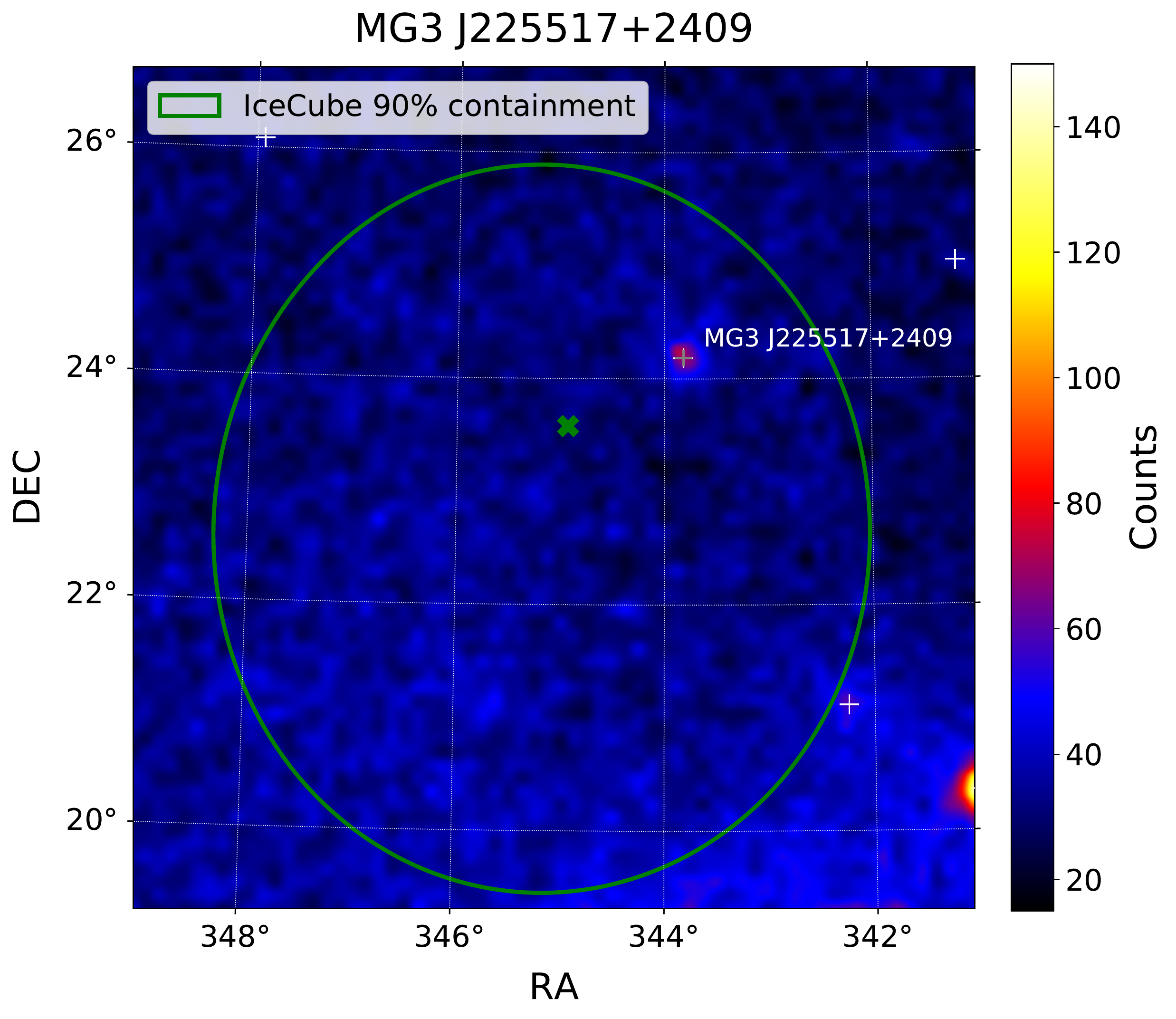}
           \includegraphics[width=8.5cm]{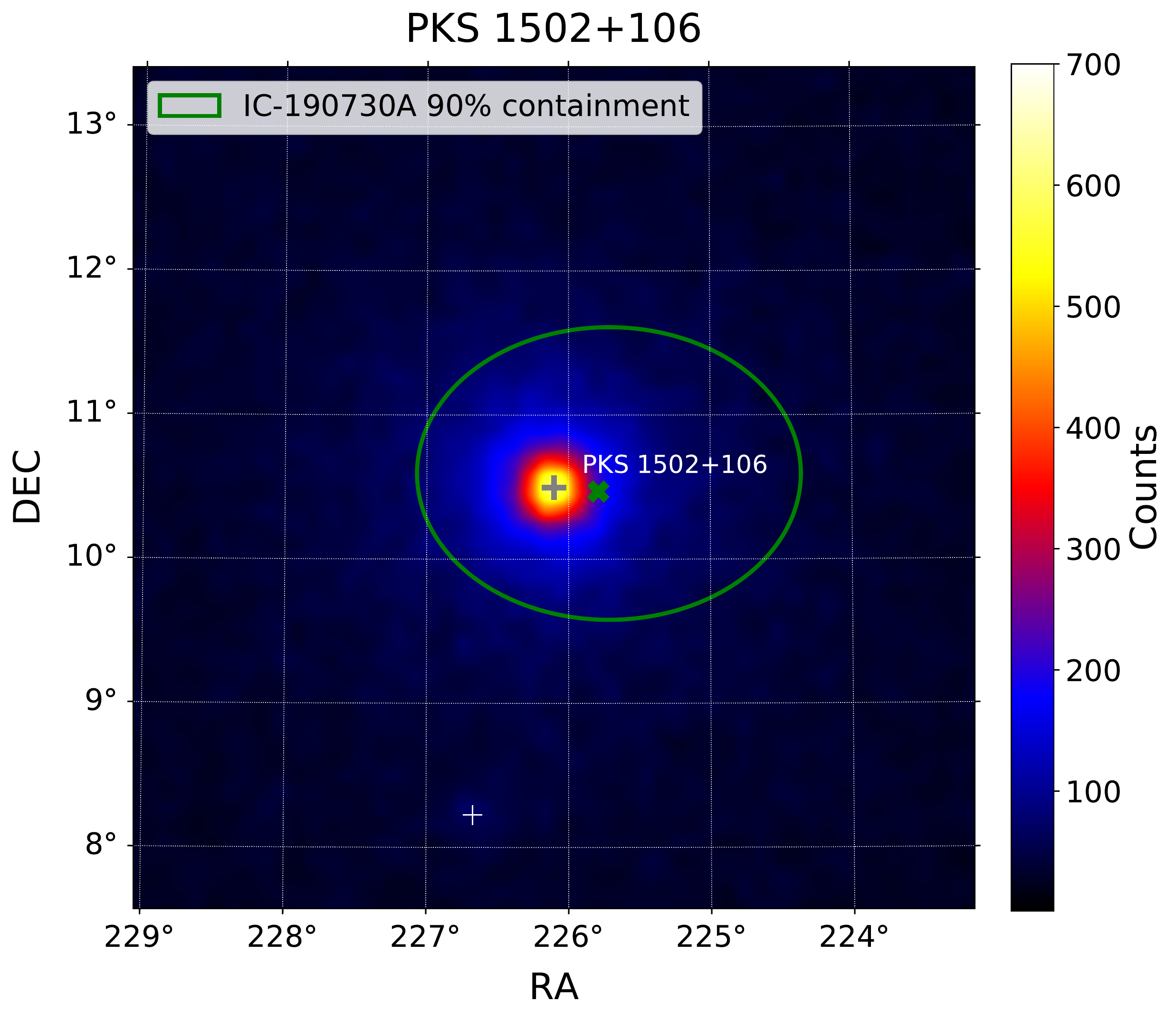}
     \caption{Gamma-ray count maps of \HH, \MG and \PKS integrated over 11 years of \textit{Fermi}-LAT data. The green cross and green line show the best-fit neutrino position 90\% uncertainty respectively. White crosses are 4FGL sources included in the background model. The count maps cover the energy range of 100\,MeV to 800\,GeV, except of \HH where we start at 1\,GeV to suppress the significant Galactic diffuse emission at the source's latitude of $b=-18.7$. The \HH count maps is not overlaid with a neutrino contour since it was identified in the neutrino flare search from 3LAC sources, i.e. the neutrino flare candidate is by definition located at the position of \HH.
              }
         \label{fig:countsmaps}
\end{figure*}

The radio-loud narrow line Seyfert 1 galaxy \HH at $z=0.061$ \citep[][]{2007ApJ...658L..13Z,2009ApJ...707L.142A} shows increased gamma-ray activity during the reported neutrino flare time (see Fig.~\ref{fig:LC_1H0323+342_zoom}). The gamma-ray countsmap integrated over 11-years of data is shown in Fig.~\ref{fig:countsmaps}. The neutrino arrived during a mild excess in gamma rays of $F_{\gamma}^{\textrm{peak}} = (2.8 \pm 0.7)\times 10^{-7}$\,ph\,cm$^{-2}$\,s$^{-1}$ and roughly one month after a flare in X-rays, UV and optical (see Fig.~\ref{fig:LC_1H0323+342_zoom}).
The chance probability to find the neutrino in a period of increased gamma-ray activity at the level of $F_{\gamma}^{\textrm{peak}}$ or higher is $p_{\gamma} = 8\%$. The neutrino flare arrives in the time bin just next to the peak. 
We note that the source shows even stronger flares at earlier times, which are not found connected to neutrino emission. 
 
Fig.~\ref{fig:SED} (upper left panel) shows the broadband SED of \HH.

The X-ray spectrum reveals a break at $\sim$3\,keV. The spectrum before the break energy is soft ($\Gamma_{1}\sim2$, see Table~\ref{tab:sed}), possibly due to coronal emission \citep[][]{2009ApJ...707L.142A,2014ApJ...789..143P,2019ApJ...872..169P}. Note that various spectral features are detected in the X-ray spectrum of this source which includes a soft X-ray excess below 2 keV, an Fe K-alpha emission line at $\sim$6 keV, and a possible Compton hump at higher frequencies \citep[see, e.g.,][for details]{2014ApJ...789..143P,2018MNRAS.475..404K,2018MNRAS.479.2464G,2019ApJ...872..169P}. Covering these aspects is beyond the scope of this work. Furthermore, the broadband SED modeling of this object suggests the gamma-ray emission region to lie well within the broad line region (BLR), i.e., close to the central black hole \citep[][]{2009ApJ...707L.142A,2014ApJ...789..143P,2018MNRAS.475..404K}. If so, the X-ray photons from the corona could constitute a target photon field for photo-hadronic interactions producing high-energy neutrinos. In particular, the interaction of the protons with the thermal continuum with a characteristic temperature ($T^*$) can produce neutrinos with energy $E_{\nu}\sim100$ TeV (T$^{*}/10^{5}$ K)$^{-1}$ \citep[see, e.g.,][]{2019ApJ...874L..29R}. The X-ray coronal photons would also absorb the gamma rays via pair-production, leading to a steepening of the gamma-ray spectrum which is observed \citep[Fig.~\ref{fig:SED};][]{2009MNRAS.397..985G,2014ApJ...789..143P,2019ApJ...874L..29R}. Another observational signature for this process is the detection of a bright X-ray emission with a soft spectral shape \citep[see,][for details]{2009MNRAS.397..985G}, which is reflected in the X-ray spectrum of \HH (Fig.~\ref{fig:SED}). A quantitative discussion will be the subject of a separate publication.
 
\begin{deluxetable*}{lcccccc}
\tablecaption{Summary of SED analysis.\label{tab:sed}}
\tablehead{
& & & \colhead{\textbf{\textit{Fermi}-LAT}} &  & &  \\
 Name & Time Window & Flux & power-law index & \multicolumn{2}{c}{log-parabola indices} & TS \\
 & MJD & $10^{-7}$\,ph cm$^{-2}$ s$^{-1}$ & & $\alpha$ & $\beta$  & }  
\startdata
 \multicolumn{7}{c}{11 years averaged} \\
 \hline
 1H 0323+342      & 54682$-$58633  &0.45~$\pm$~0.02 & -- & 2.77~$\pm$~0.04 & 0.09~$\pm$~0.04 & 1027   \\
 PKS 1502+106     & 54682$-$58695  &2.97~$\pm$~0.02 & -- & 2.12~$\pm$~0.01 & 0.10~$\pm$~0.01 & 95412  \\
 MG3 J225517+2409 & 54682$-$58633  &0.11~$\pm$~0.01 & 2.03~$\pm$~0.04 & -- & -- & 955  \\
 \hline
  \multicolumn{7}{c}{Contemporaneous} \\
   \hline
 1H 0323+342      & 57263$-$57392  &1.24~$\pm$~0.21 & -- & 3.25~$\pm$~0.47 & 0.24~$\pm$~0.25 & 90  \\
 PKS 1502+106     & 58664$-$58724  &0.86~$\pm$~0.22 & -- & 2.31~$\pm$~0.16 & 0.01~$\pm$~0.07 & 97  \\
 MG3 J225517+2409 & 55346$-$55501 &0.41~$\pm$~0.08 & 2.02~$\pm$~0.09 & -- & -- & 181 \\
  \hline
 \multicolumn{7}{c}{Interesting multi-wavelength features of \PKS} \\
  \hline
 PKS 1502+106     & 54682$-$54692  &19.48~$\pm$~0.81 & -- & 1.87~$\pm$~0.04 & 0.12~$\pm$~0.02 & 4205 \\
                  & 55266$-$57022 &0.88~$\pm$~0.03 & -- & 2.27~$\pm$~0.02 & 0.07~$\pm$~0.01 & 56318 \\
                  & 57210$-$57219 &14.15~$\pm$~0.69 & -- & 1.74~$\pm$~0.05 & 0.09~$\pm$~0.02 & 4830 \\
                  & 58107$-$58125   &4.80~$\pm$~0.35 & -- & 2.16~$\pm$~0.06 & 0.06~$\pm$~0.04 & 25676 \\
 \hline
  \hline
 \multicolumn{7}{c}{\textbf{\textit{Swift}-XRT}}  \\
 Name & Exposure & $\Gamma_{1}$ & $\Gamma_{2}$ & Flux & Normalization & $\chi^{2}$/dof\\
 & ks & & & 10$^{-12}$ erg cm$^{-2}$ s$^{-1}$ & 10$^{-4}$ ph cm$^{-2}$ s$^{-1}$ keV$^{-1}$ & \\
\hline
\multicolumn{7}{c}{Contemporaneous} \\
 \hline
 1H 0323+342      & 32.81  & 2.04$^{+0.04}_{-0.04}$ & 1.67$^{+0.10}_{-0.13}$ & 22.94$^{+0.57}_{-0.64}$ & 37.50$^{+0.70}_{-0.69}$ & 366.91/285 \\
 PKS 1502+106     & 5.61 & 1.00$^{+0.30}_{-0.31}$ &               & 1.12$^{+0.41}_{-0.28}$ & 0.67$^{+0.21}_{-0.18}$ & 56.59/66\\
 MG3 J225517+2409 & 1.65 & 2.03$^{+1.00}_{-1.06}$ &               & 0.50$^{+0.53}_{-0.20}$ & 0.90$^{+0.45}_{-0.34}$ & 12.57/12\\
  \hline
  \multicolumn{7}{c}{Interesting multi-wavelength features of \PKS} \\
   \hline
 PKS 1502+106     & 36.67 & 1.44$^{+0.06}_{-0.06}$ &               & 2.16$^{+0.14}_{-0.12}$ & 2.44$^{+0.12}_{-0.13}$ & 63.62/69\\
                  & 13.50 & 1.58$^{+0.25}_{-0.25}$ &               & 0.73$^{+0.17}_{-0.13}$ & 0.94$^{+0.14}_{-0.15}$ & 12.51/7\\
                  & 11.06 & 1.69$^{+0.17}_{-0.16}$ &               & 1.22$^{+0.17}_{-0.15}$ & 1.75$^{+0.21}_{-0.21}$ & 20.60/12\\
                  & 14.69 & 1.57$^{+0.10}_{-0.10}$ &               & 2.12$^{+0.21}_{-0.20}$ & 2.72$^{+0.21}_{-0.21}$ & 42.60/29\\
 \hline
  \hline
 \multicolumn{7}{c}{\textbf{\textit{Swift}-UVOT}} \\
 Name & V & B & U & W1 & M2 & W2\\
\hline
\multicolumn{7}{c}{Contemporaneous} \\
 \hline
 1H 0323+342  & 19.58~$\pm$~0.23 & 19.85~$\pm$~0.19 & 22.82~$\pm$~0.26 & 21.58~$\pm$~0.30 & 24.94~$\pm$~0.38 & 23.47~$\pm$~0.31\\
 PKS 1502+106 &                  &                  & 0.90~$\pm$~0.06  &                  & 0.67~$\pm$~0.06  &                 \\
 MG3 J225517+2409 & 2.91~$\pm$~0.32 & 1.90~$\pm$~0.13& 2.26~$\pm$~0.12& 1.92~$\pm$~0.12& 2.74~$\pm$~0.25& 1.54~$\pm$~0.10\\
  \hline
  \multicolumn{7}{c}{Interesting multi-wavelength features of \PKS} \\
   \hline
 PKS 1502+106 & 4.71~$\pm$~0.11  & 4.45~$\pm$~0.08  & 3.97~$\pm$~0.08  & 2.59~$\pm$~0.06  & 2.67~$\pm$~0.07  & 2.21~$\pm$~0.06 \\
              & 0.55~$\pm$~0.06  & 0.67~$\pm$~0.05  & 0.61~$\pm$~0.06  & 0.55~$\pm$~0.05  & 0.51~$\pm$~0.04  & 0.43~$\pm$~0.03 \\
              & 6.38~$\pm$~0.24  & 6.24~$\pm$~0.18  & 5.32~$\pm$~0.16  & 3.71~$\pm$~0.14  & 3.57~$\pm$~0.11  & 2.98~$\pm$~0.11 \\
              & 8.06~$\pm$~0.20  & 7.20~$\pm$~0.15  & 6.42~$\pm$~0.15  & 5.03~$\pm$~0.16  & 4.68~$\pm$~0.13  & 4.34~$\pm$~0.10 \\
    \enddata
\tablecomments{The first block shows the results of the \textit{Fermi}-LAT SED results performed in a given time window. The quoted gamma-ray flux is integrated in the 0.1$-$800 GeV energy range. $N_{pred}$ is the number of predicted gamma-ray photons in the given time window. The SED is modeled with a power-law, unless a log-parabola description with spectral parameters $\alpha$ and $\beta$ results in a significantly better fit to the data. The second block shows the results of the \textit{Swift}-XRT spectral analysis. $\Gamma_1$ is the photon index of a power-law model or photon index before the break energy in a broken power-law model, while $\Gamma_2$ is the photon index after the break energy in the broken power-law model. The flux is integrated in the 0.3$-$10 keV energy range and the normalization is defined at 1 keV in units of 10$^{-4}$ ph cm$^{-2}$ s$^{-1}$ keV$^{-1}$. Absorption by the Galactic neutral hydrogen is taken into account using the following column densities along the line of sight \citep[][]{2005AnA...440..775K}: $N_{\rm H} = 1.17\times10^{21}$ cm$^{-2}$ (1H 0323+342), $2.03\times10^{20}$ cm$^{-2}$ (PKS 1502+106), and $3.57\times10^{20}$ cm$^{-2}$ (MG3 J225517+2409). The third block shows the results of the \textit{Swift}-UVOT analysis and gives the average flux in the {\it Swift} V, B, U, W1, M2 and W2 bands in the units of 10$^{-12}$ erg cm$^{-2}$ s$^{-1}$. The Swift analyses were performed in the same time periods specified for the Fermi-LAT results.}
\end{deluxetable*}

\subsubsection{\MG}

The distant BL Lac object \MG shows a major flare coincident with the neutrino arrival time (see Fig.~\ref{fig:LC_MG3J225517+2409}). 4LAC reports a redshift of 1.37~\citep{Fermi-LAT:2019pir}, which is taken from SDSS. However, the extracted redshift is flagged as ``chi-squared of best fit is too close to that of second best ($< 0.01$ in reduced chi-squared)''. \citet{PaianoATel} find that the redshift is $>0.8633$. The gamma-ray flare reaches a flux level of $(3.5\pm1.0)\times 10^{-8}$\,ph\,cm$^{-2}$\,s$^{-1}$ and lasts roughly 140\,days (see Fig.~\ref{fig:LC_MG3J225517+2409}). The chance probability to find the neutrino in a period of increased gamma-ray activity at the this level or higher is $p_{\gamma} = 4\%$.

Fig.~\ref{fig:SED} (upper right panel) shows the broad band SED of \MG and the best-fit spectral values for the gamma-ray, X-ray and UV bands are provided in Table~\ref{tab:sed}.

\subsubsection{\PKS}

The flat-spectrum radio quasar (FSRQ) \PKS was found to be located within the 50\% uncertainty region of IC-190730A. 
The neutrino was reported with a signalness of $67\%$ and an energy of 300\,TeV~\citep{GCNPKS}. \PKS is the 15th brightest out of 2863 source in the 4LAC catalog in terms of gamma-ray energy flux at $>100$\,MeV despite its large redshift of 1.84~\citep{2010MNRAS.405.2302H}, suggesting an extremely high intrinsic luminosity.

\begin{figure*}
    \centering
           \includegraphics[width=8.5cm]{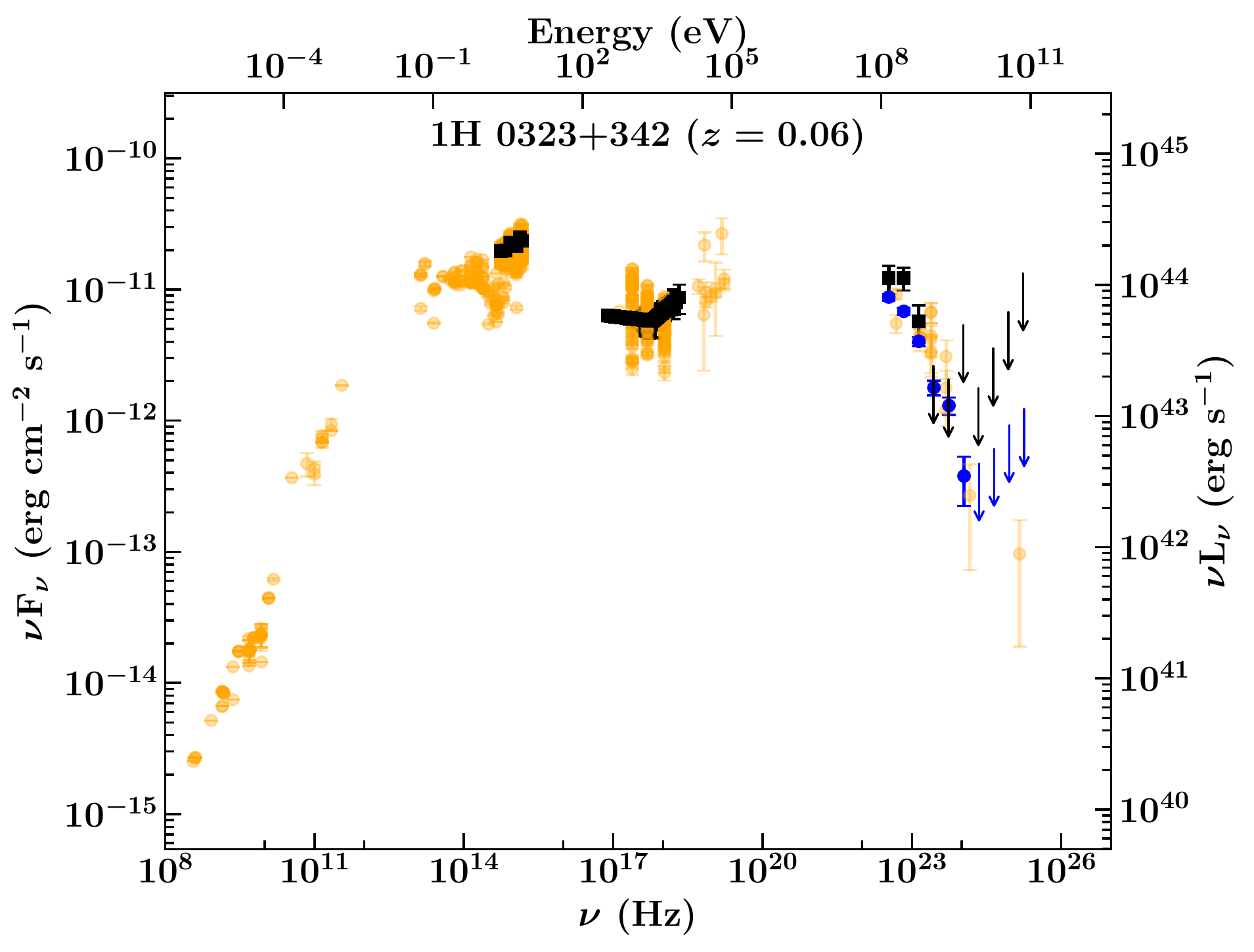}
           \includegraphics[width=8.5cm]{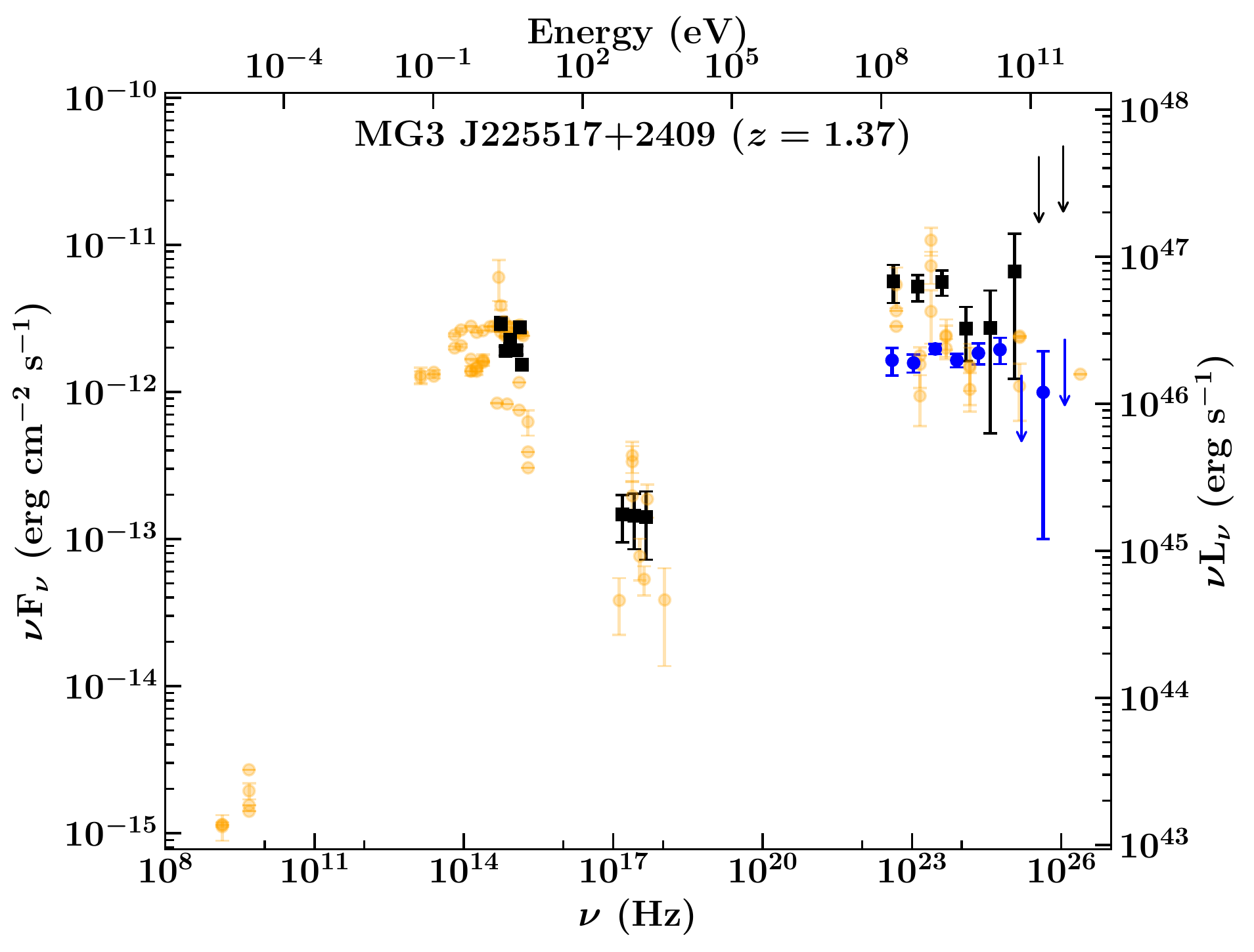}
           \includegraphics[width=8.5cm]{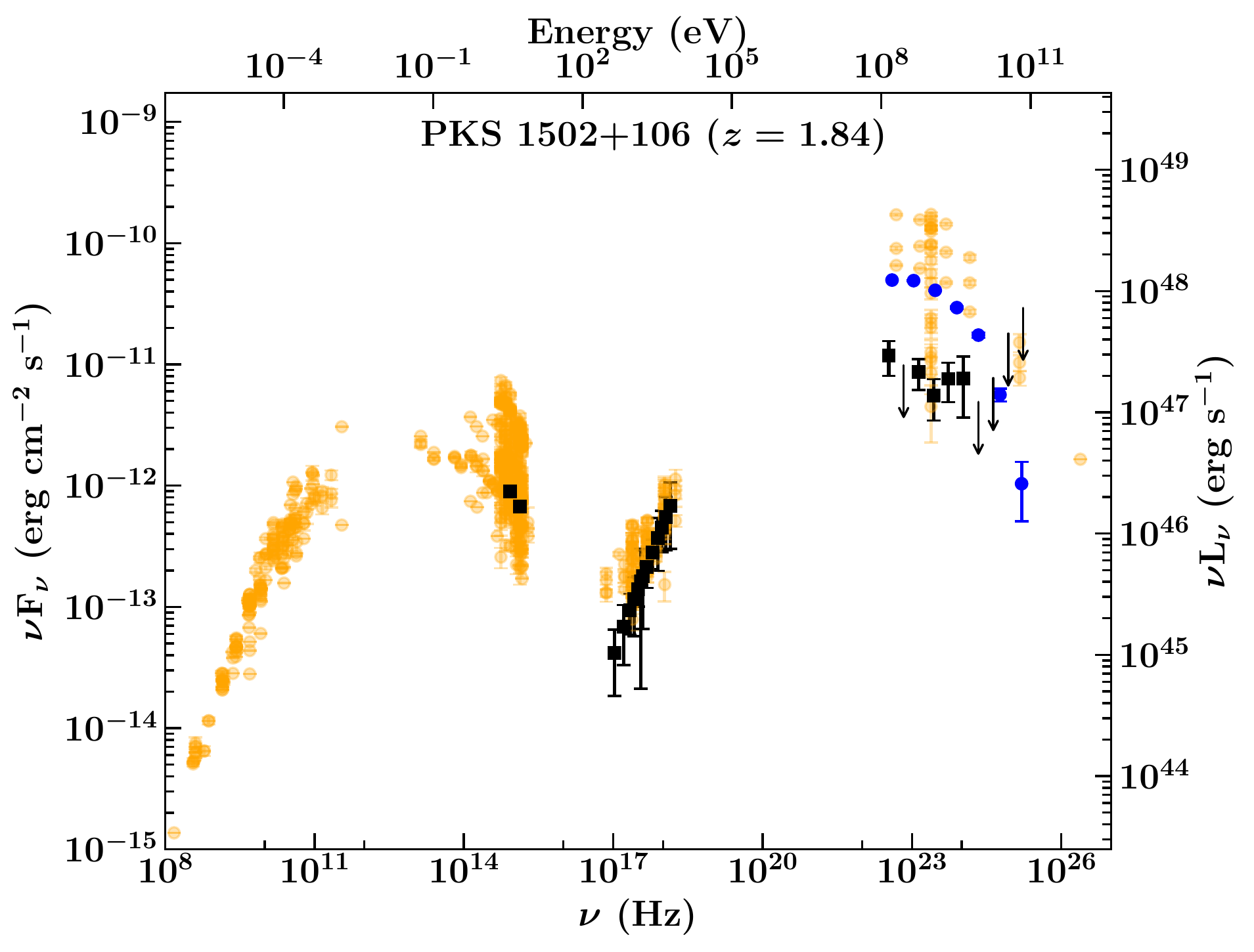}
           \includegraphics[width=8.5cm]{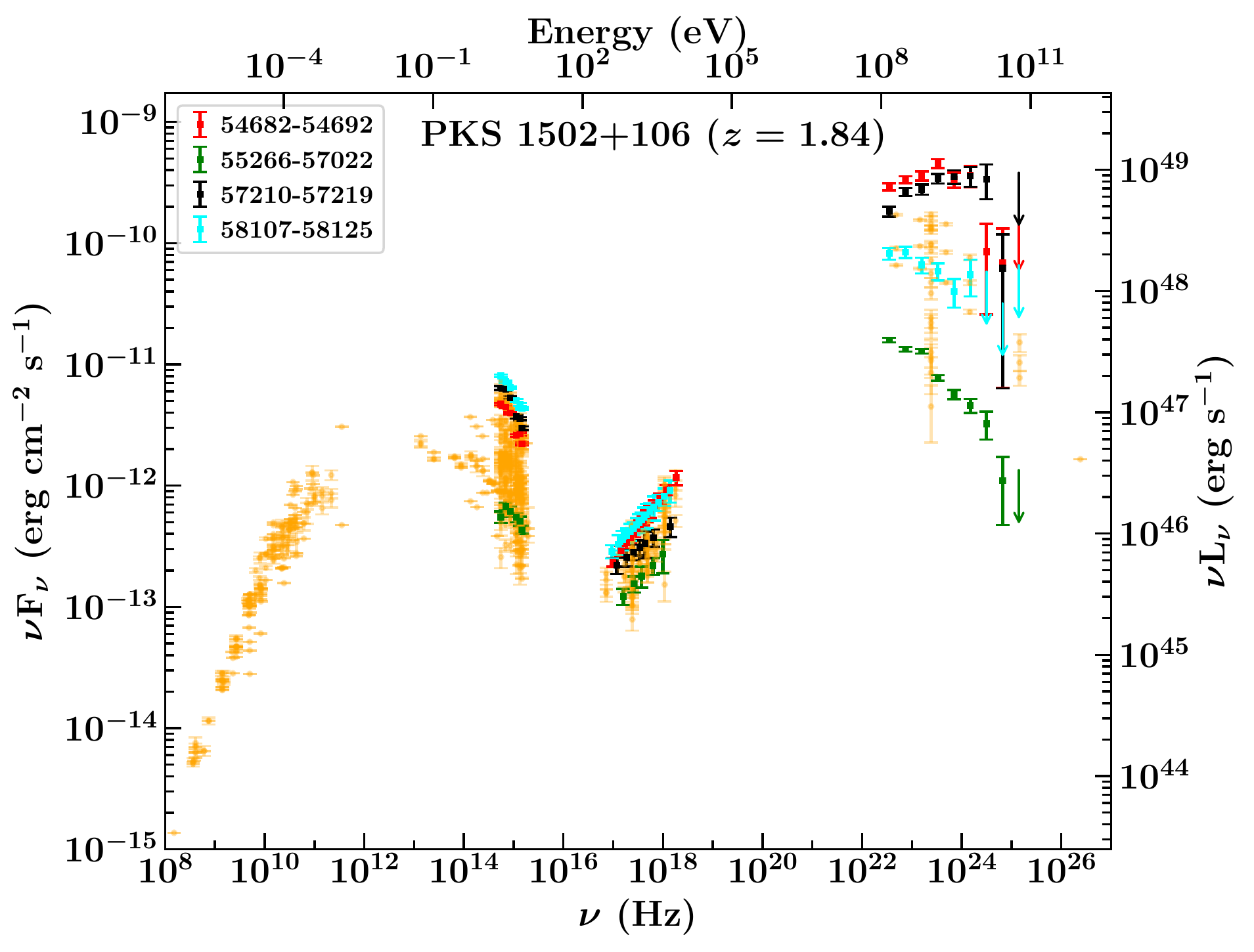}
     \caption{Spectral energy distribution of \HH, \MG and \PKS. Archival data are shown in orange. The 11-year {\it Fermi}-LAT SEDs are overlayed in blue circles. The black data points refer to the observations contemporaneous to the epoch of neutrino detection. The lower right panel shows the SEDs of \PKS during 4 selected epochs (see Tab.~\ref{tab:sed} for details).
              }
         \label{fig:SED}
\end{figure*}

It was found to be in a low activity state during the arrival time of the high-energy neutrino (see Fig.~\ref{fig:LC_PKS1502+106} and \ref{fig:SED}). However, the OVRO radio light curve of \PKS shows a long-term outburst starting in 2014 and reaching the highest flux density ever reported from this source (since the beginning of the OVRO measurements in 2008) during the arrival of the $300$\,TeV neutrino IC-190730A \citep{2019ATel12996....1K}. \TXS showed a similar increase in the radio emission observed by OVRO in coincidence with IC-170922A~\citep{MWScience,2019ATel12996....1K}. A strong increase in radio emission was also determined in VLBI data for another blazar PKS B1424-418, which was found coincident with a high-energy but poorly reconstructed neutrino event, by \citet{2016NatPh..12..807K} and \citet{2020arXiv200100930P} find a correlation of IceCube neutrinos with radio-bright AGN with a 0.2\% p-value. Quantifying the chance coincidence of a radio flare with the arrival time of a neutrino is out of the scope of this paper.

Fig.~\ref{fig:SED} (lower left panel) shows the broadband SED of \PKS. The 11 years averaged gamma-ray spectrum of this source reveals a significant curvature/break which could be reflecting the shape of the particle spectrum or be due to extrinsic absorption by the BLR photons. Interestingly, the EBL absorption is not significant below 50 GeV at $z=1.84$ \citep[][]{2018Sci...362.1031F} and the gamma-ray emission from PKS 1502+106 has been explained by the interaction of the jet electrons with the BLR photons \citep[e.g.,][]{2010ApJ...710..810A}. Therefore, the observed spectral curvature could be due to gamma-ray absorption by the BLR photons via the pair production process and/or transition from the Thomson to Klein-Nishina regime. If so, the same BLR photon field could also act as a target photon field for neutrino production by interacting with the hadrons present in the jet \citep[see, e.g.,][]{2019ApJ...874L..29R}.

The gamma-ray spectral index shows variation in time (see Fig.~\ref{fig:LC_PKS1502+106}, second panel). The spectrum tends to harden when the gamma-ray flux increases. The hard spectral regions indicate an increase in high-energy emission and are therefore promising targets for follow-up searches of $\mathcal{O}$(TeV) neutrinos. 
Since \PKS is the most interesting source of our sample (due to its high gamma-ray energy flux), we study the spectral behavior during the multi-wavelength flares in more detail to give guidance for future neutrino searches. We split the 11-yr light curve in four regions of interest, where we obtain the gamma-ray spectral shape (see Tab.~\ref{tab:sed}). We select one period from MJD 55266-57022 to cover the quiet state in gamma rays and three short periods of roughly 10-day length chosen to cover the three bright X-rays flares, which are also accompanied by optical flares (see Fig.~\ref{fig:LC_PKS1502+106}).
We find that during the gamma-ray quiet state the flux values in each wavelength reach a minimum flux level (shown in green). Interestingly, the highest flare in X-ray and optical (cyan) does not correspond to the highest flare in gamma rays, while the highest gamma-ray activity is also accompanied by a significant increase in optical and X-rays. The different flaring behavior indicates different conditions of the emission region in the source. Detailed time-dependent modeling, which is outside of the scope of this work, could give a deeper insight into the variable nature of the source. 

\subsection{Source Population}

\begin{figure*}
    \centering
           \includegraphics[width=8.5cm]{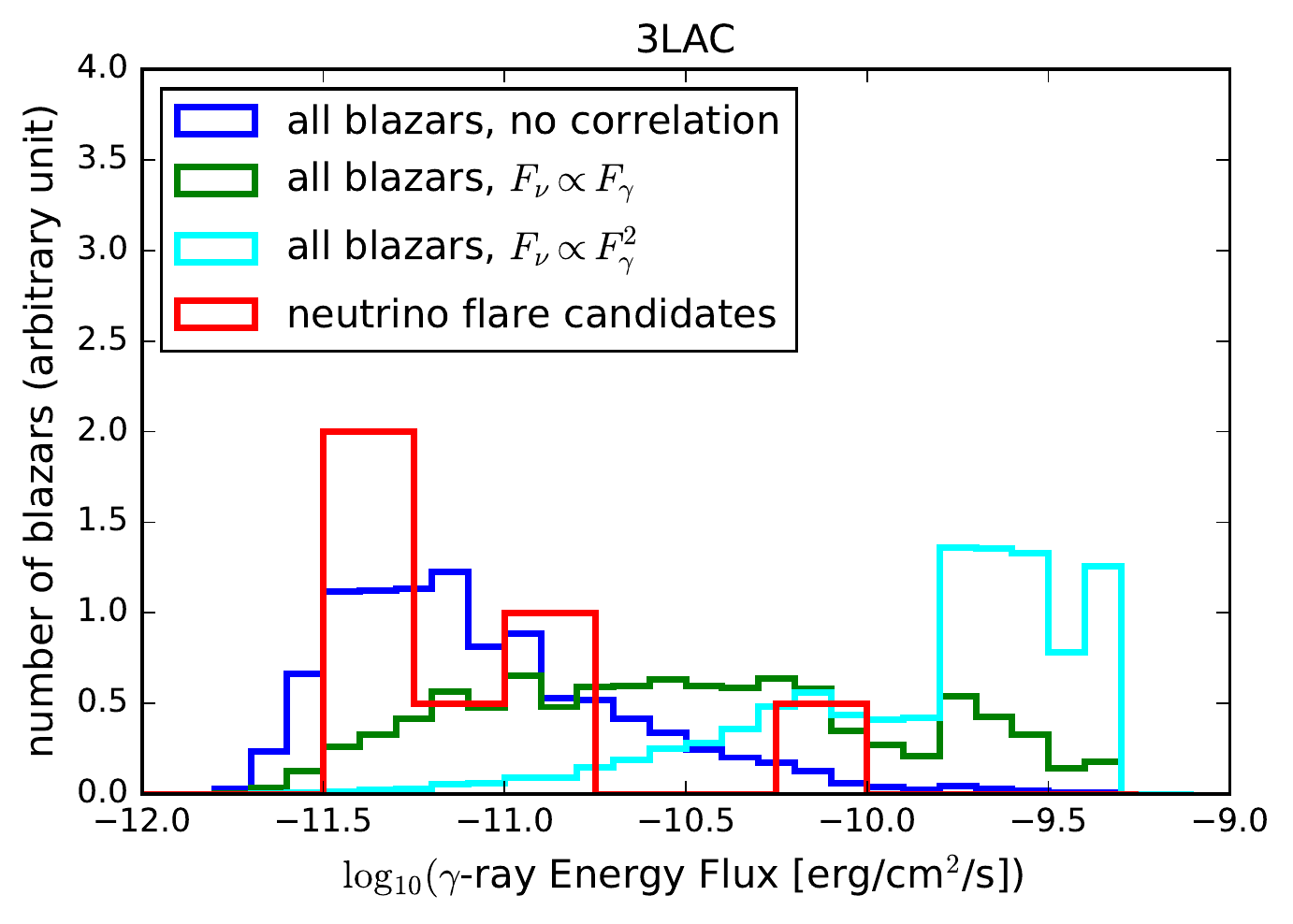}
          \includegraphics[width=8.5cm]{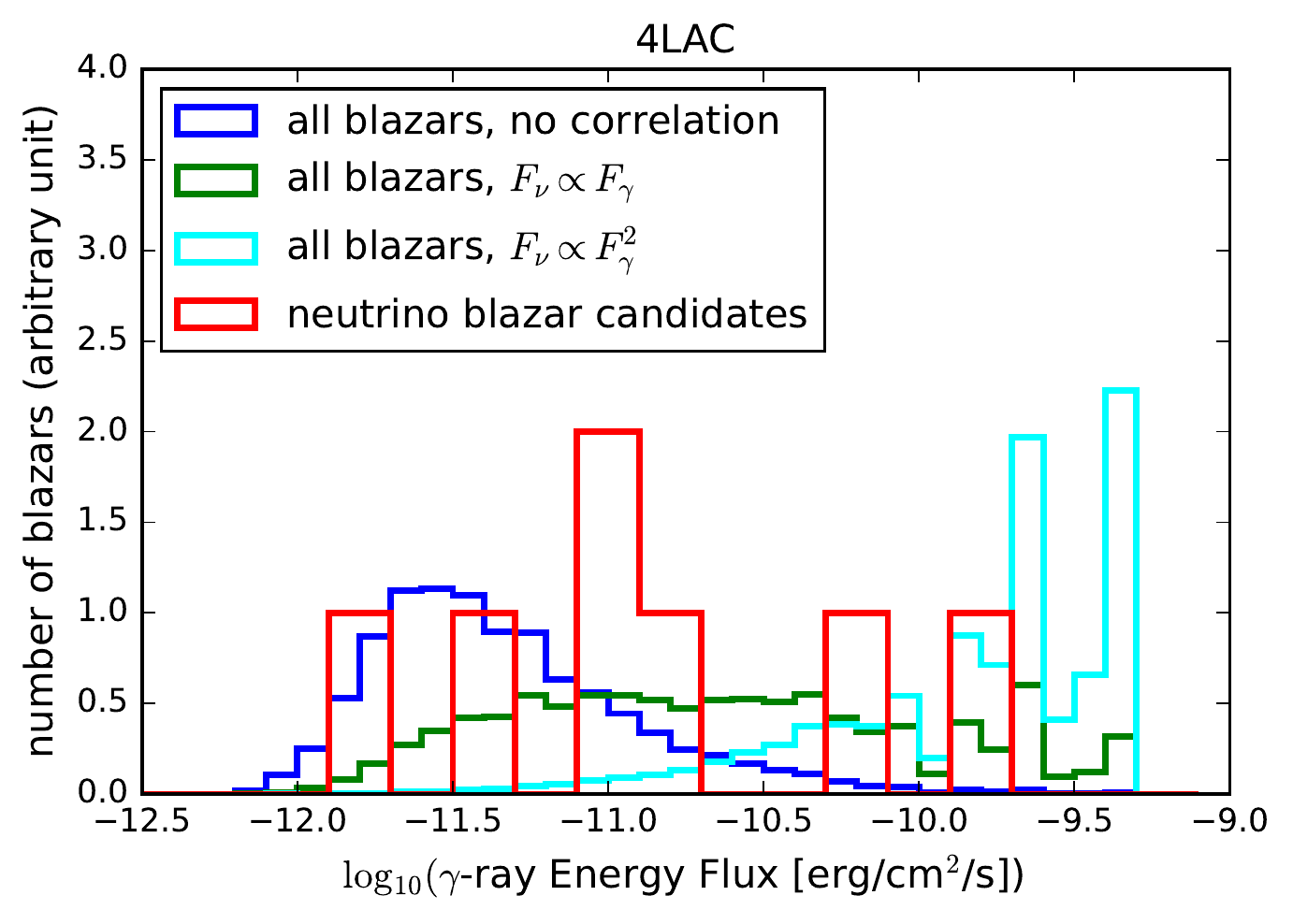}

     \caption{Expected gamma-ray energy flux (in the 100\,MeV to 100\,GeV energy range) distribution of all blazars in 3LAC (left) and 4LAC (right) according to the no-correlation hypothesis shown in blue. The expected distribution according to the hypothesis of a linear (quadratical) correlation between neutrino and gamma-ray energy flux is shown in green (cyan). Left: Observed gamma-ray energy flux of neutrino flare candidate blazars from 3LAC overlaid in red. The red distribution matches the blue one, but not the green or cyan. Right: Observed gamma-ray energy flux of 4LAC blazars found in spatial coincidence with single high-energy neutrinos (red). 
              }
         \label{fig:KS}
\end{figure*}

As a result of the KS test (see Tab.~\ref{table:KS}), we find that the blazars found in coincidence with single high-energy neutrinos are well described by both the gamma-ray energy flux distribution expected in case of a linear correlation between neutrino and gamma-ray energy flux and the hypothesis of no correlation between the two fluxes. 
If all sources are combined, the single neutrino source candidates are compatible with the no-correlation hypothesis with a p-value of $12.6\%$ and consistent with the expectation in case of a linear correlation between neutrino and gamma-ray energy flux (p-value of $64\%$). They show a mismatch with the hypothesis of a quadratical correlation (p-value of $0.03\%$). The gamma-ray energy flux distribution is illustrated in Fig.~\ref{fig:KS}. 

\MG would have failed our angular uncertainty requirement selection and was not included in the KS test.

At the same time the candidate neutrino-flare sources show a good match with the random distribution (p-value of $39-98\%$), but are less well described by the energy-flux weighted distribution (p-value of $0.4-2.1\%$).

\section{Summary and Conclusions}
\label{sec:Conclusion}
 
\subsection{Individual sources} 
In summary, the available data do not show any significant temporal correlation with the neutrino arrival time considering all blazars studied here, which would allow us to identify one of the sources as potential cosmic-ray acceleration site. This is consistent with findings by \citet{Righi:2018hhu}, who studied the gamma-ray light curves of 7 BL Lac objects and did not find a clear pattern in common among the sources.

Most sources are well observed in GeV gamma rays and optical wavelengths, where no significant temporal correlation with the neutrino emission is found. For the sources monitored by OVRO, no short-term features related to the neutrino arrival time are observed. 
Three out of five sources coincident with single high-energy neutrinos are monitored by OVRO in radio and two (\TXS and \PKS) show a long-term increase of the radio flux density, which peaks during the neutrino arrival time. The third one (\MG) is only covered by OVRO observations 70\,days after the neutrino arrival time, but might be compatible with a radio flux increase assuming a smooth extrapolation of the temporal behavior to earlier times.
The five neutrino flare source candidates, which are monitored in radio, show no correlation with long-term radio activity. Radio monitoring of future neutrino blazar coincidences could reveal if there is indeed a connection between the radio and single high-energy neutrino emission.

\subsection{Source Population}

We find that the single high-energy neutrino coincidences with blazars are consistent with a p-value of $12.6\%$ with being due to random chance. At the same time they are consistent with the hypothesis that the single high-energy neutrino emission is correlated linearly with the gamma-ray brightness of the blazars. This interpretation is consistent with the association of IC-170922A and the bright gamma-ray flare of \TXS.

We note that the contribution of gamma-ray blazars to the diffuse neutrino flux is constrained by blazar stacking analyses~\citep{Aartsen:2016lir}, which constrain the blazar contribution to less than $27\%$ under the assumption that the neutrino spectrum follows an unbroken $E^{-2.5}$ power law. Assuming a steeper power-law with index of 2, which is compatible with the diffuse flux fit above $\sim200$\,TeV weakens the constrains on the diffuse flux contribution to 40-80\%~\citep{ICScience} and leaves room for a significant contribution from gamma-ray blazars. \citet{2018A&A...620A.174K} studied the gamma-ray and X-ray emission of 3LAC blazars in the vicinity of high-energy starting events detected by IceCube and find no direct correlation between the \textit{Fermi}-LAT gamma-ray flux and the IceCube neutrino flux.

Assuming that the observed linear correlation of single high-energy neutrinos and the average gamma-ray energy flux is genuine, our results would have many broad implications. It would allow us to utilize neutrino blazar coincidences for the study of cosmic-ray acceleration. Furthermore, it would allow for an effective search for more coincidences to further characterize the population of sources of high-energy neutrinos.

At the same time the candidate neutrino-flare emitting blazars are compatible with the background hypothesis and the data do not support the hypothesis that neutrino emission is correlated to the average gamma-ray energy flux. This could indicate that either neutrino flares are not accompanied by strong gamma-ray emission, or that these coincidences are of a random nature. The first case could be realized in sources where neutrinos are produced in regions optically thick to gamma rays, where gamma-ray emission is absorbed (so-called hidden sources) and cascades to the X-ray band, where we do not have good observational constraints from archival data~\citep[see also][]{Gao:2018mnu,Keivani:2018rnh}.

\begin{acknowledgements}

This work was supported by the Initiative and Networking Fund of the Helmholtz Association. We thank Xavier Rodrigues, Shan Gao and Walter Winter for fruitful discussions. 
B.J.S. is supported by NSF grants AST-1908952, AST-1920392, and AST1911074.
S.K. acknowledges support from the European Research Council under the European Union's Horizon 2020 research and innovation program, under grant agreement No\,771282. W.M. acknowledges support from CONICYT project Basal AFB-170002. T.H. was supported by the Academy of Finland projects 317383 and 320085.

\textbf{\textit{Fermi}-LAT:} The \textit{Fermi} LAT Collaboration acknowledges generous ongoing support
from a number of agencies and institutes that have supported both the
development and the operation of the LAT as well as scientific data analysis.
These include the National Aeronautics and Space Administration and the
Department of Energy in the United States, the Commissariat \`a l'Energie Atomique
and the Centre National de la Recherche Scientifique / Institut National de Physique
Nucl\'eaire et de Physique des Particules in France, the Agenzia Spaziale Italiana
and the Istituto Nazionale di Fisica Nucleare in Italy, the Ministry of Education,
Culture, Sports, Science and Technology (MEXT), High Energy Accelerator Research
Organization (KEK) and Japan Aerospace Exploration Agency (JAXA) in Japan, and
the K.~A.~Wallenberg Foundation, the Swedish Research Council and the
Swedish National Space Board in Sweden.
 
Additional support for science analysis during the operations phase is gratefully
acknowledged from the Istituto Nazionale di Astrofisica in Italy and the Centre
National d'\'Etudes Spatiales in France. This work performed in part under DOE
Contract DE-AC02-76SF00515.

\textbf{ASAS-SN:} ASAS-SN is supported by the Gordon and Betty Moore Foundation through grant GBMF5490 to the Ohio State University, and NSF grants AST-1515927 and AST-1908570. Development of ASAS-SN has been supported by NSF grant AST-0908816, the Mt. Cuba Astronomical Foundation, the Center for Cosmology and AstroParticle Physics at the Ohio State University, the Chinese Academy of Sciences South America Center for Astronomy (CAS- SACA), the Villum Foundation, and George Skestos.

\textbf{Others:} This research has made use of data from the OVRO 40-m monitoring program (Richards, J. L. et al. 2011, ApJS, 194, 29) which is supported in part by NASA grants NNX08AW31G, NNX11A043G, and NNX14AQ89G and NSF grants AST-0808050 and AST-1109911. 

The CSS survey is funded by the National Aeronautics and Space
Administration under Grant No. NNG05GF22G issued through the Science
Mission Directorate Near-Earth Objects Observations Program.  The CRTS
survey is supported by the U.S.~National Science Foundation under
grants AST-0909182 and AST-1313422.

Part of this work is based on archival data,
software, or online services provided by the Space Science Data
Center (SSDC).

\end{acknowledgements}

\software{Fermi-LAT Science Tools (v11r04p00), fermipy \citep[v0.17.4;][]{Wood:2017yyb}, astropy \citep{2013A&A...558A..33A,2018AJ....156..123A}, IRAF \citep{1986SPIE..627..733T,1993ASPC...52..173T}, ISIS \citep{alard98, alard00}, XSPEC \citep[][]{1996ASPC..101...17A}, HEASoft (v 6.26.1)}

\bibliography{my-bib.bib} 

\bibliographystyle{aasjournal}

\clearpage

\begin{appendix}

\section{Multi-wavelength Light Curve Plots}

\begin{figure*}[h!]
    \centering
           \includegraphics[width=18cm]{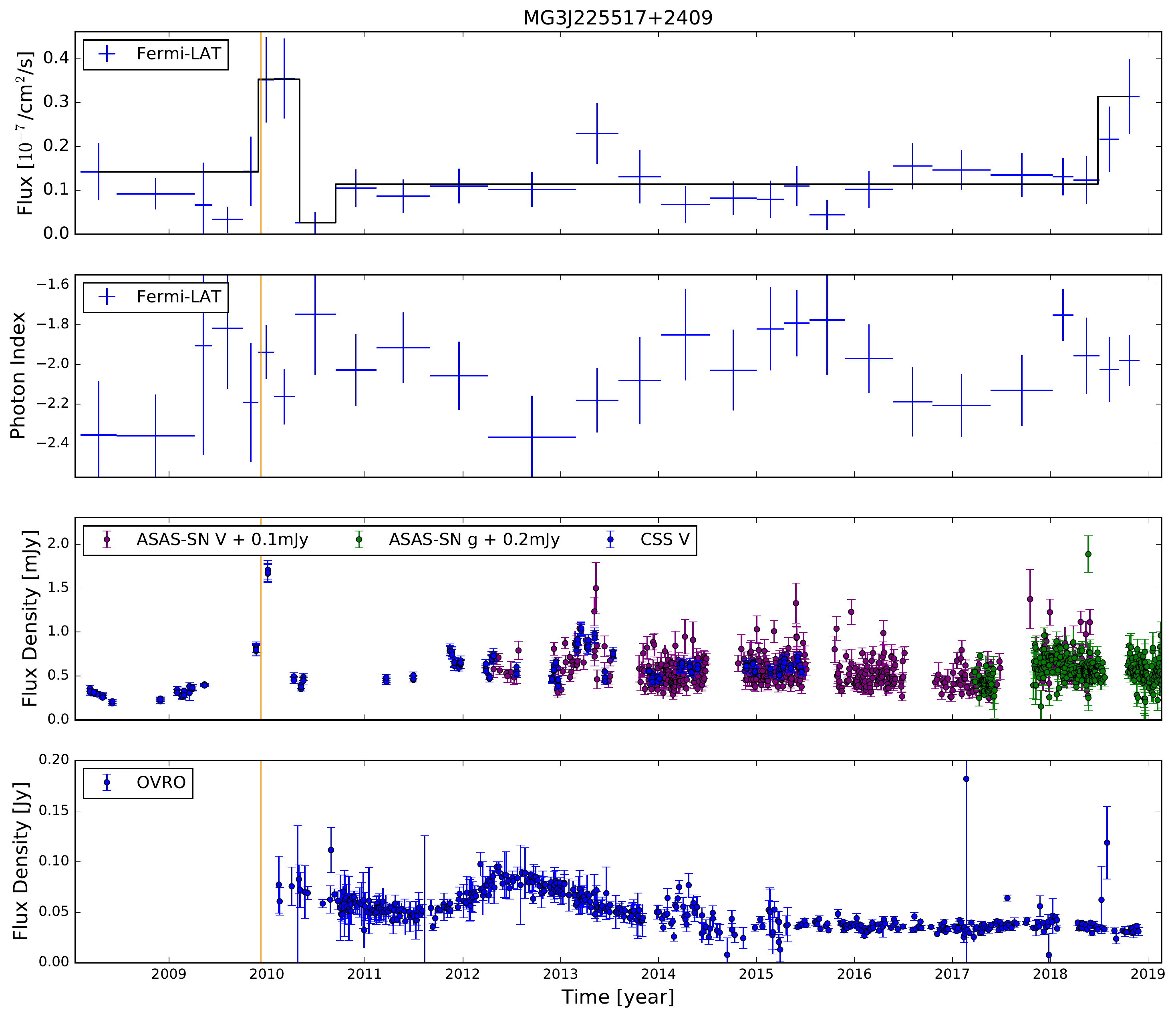}
     \caption{Multi-wavelength light curve of \MG. The orange line indicates the arrival time of the high-energy neutrino IC-100608A.
              }
         \label{fig:LC_MG3J225517+2409}
\end{figure*} 

\begin{figure*}[h!]
    \centering
           \includegraphics[width=18cm]{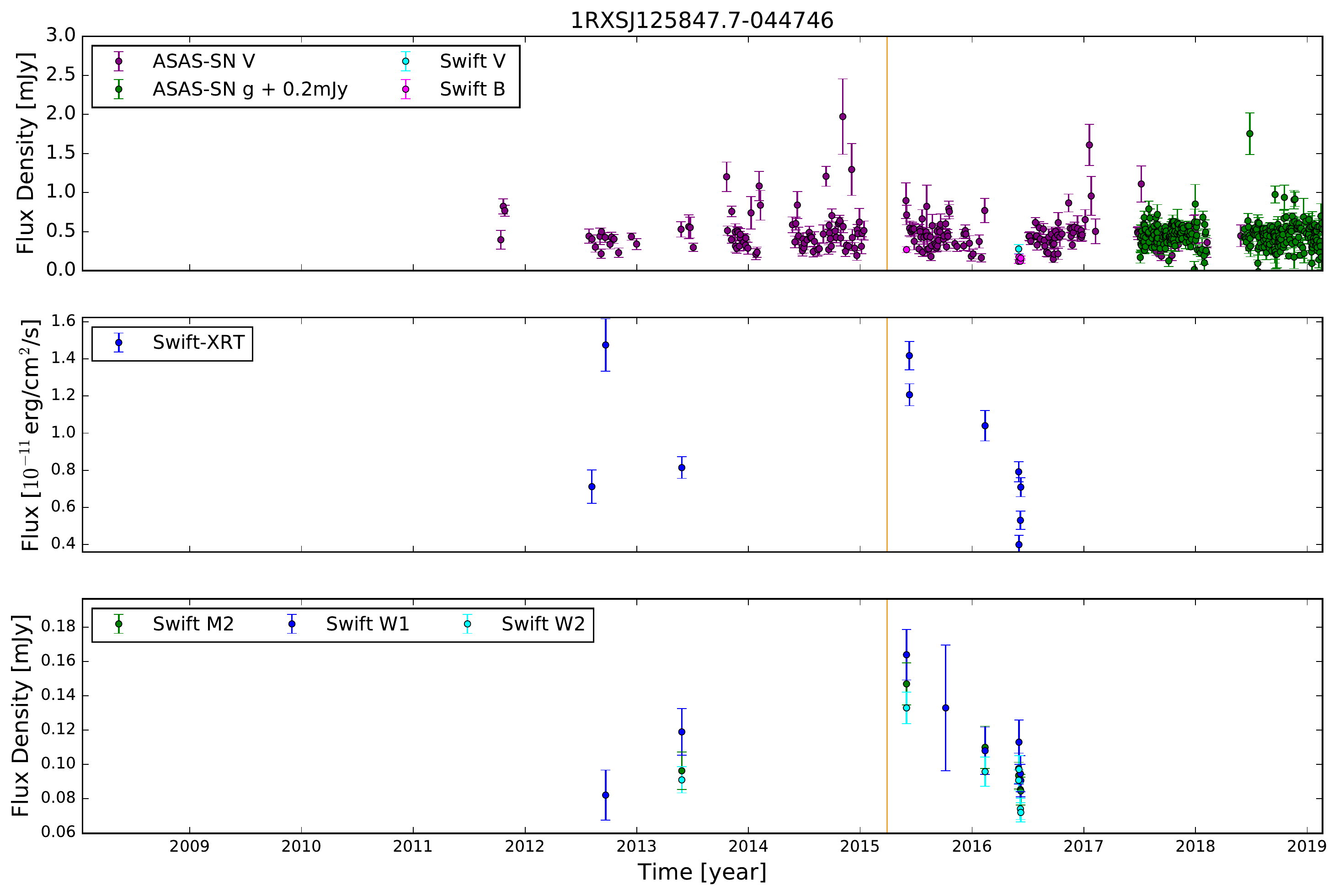}
     \caption{Multi-wavelength light curve of 1RXS J125847.7-044746. The orange line indicates the arrival time of the high-energy neutrino IC-150926A. The source is too dim in gamma rays to resolve the temporal variability.
             }
         \label{fig:LC_1RXSJ125847.7-044746}
\end{figure*} 

\begin{figure*}[h!]
    \centering
           \includegraphics[width=18cm]{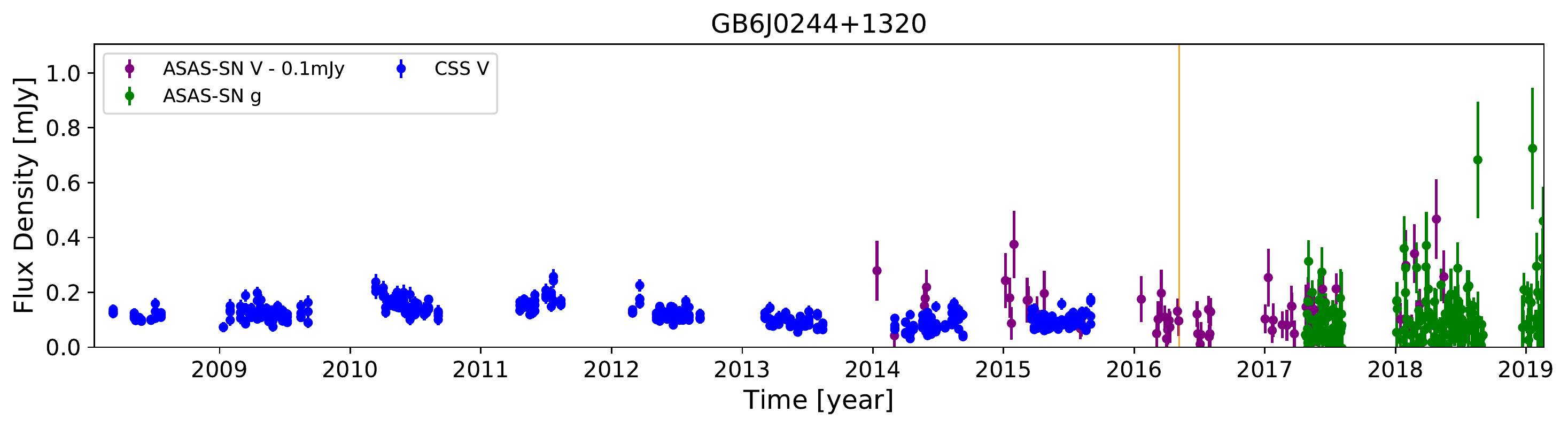}
     \caption{Multi-wavelength light curve of GB6 J0244+1320. The orange line indicates the arrival time of the high-energy neutrino IC-161103A. The source is too dim in gamma rays to resolve the temporal variability.
              }
         \label{fig:LC_GB6J0244+1320}
\end{figure*} 

\begin{figure*}
    \centering
           \includegraphics[width=18cm]{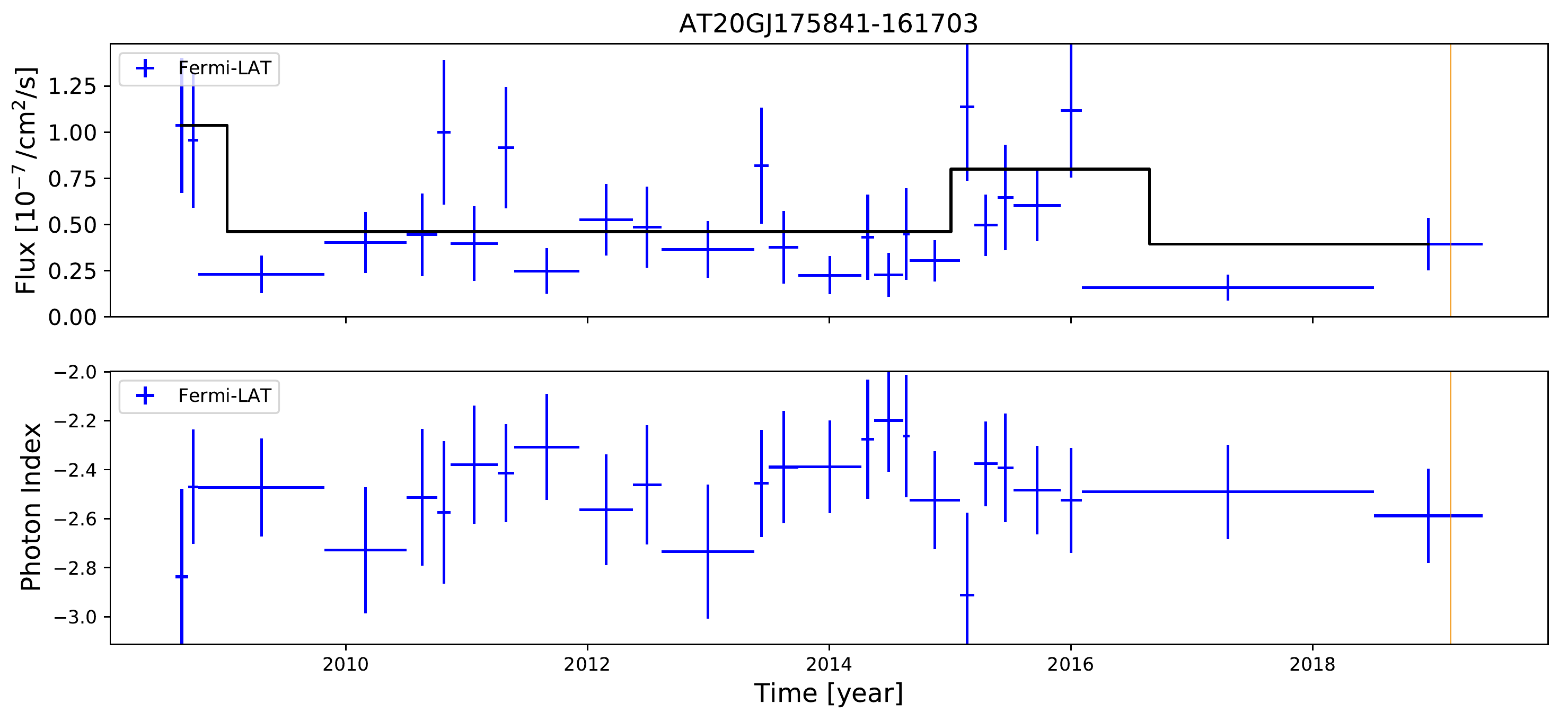}
     \caption{Multi-wavelength light curve of \AT. The orange line indicates the arrival time of the high-energy neutrino IC-190221A.
              }
         \label{fig:LC_AT20GJ175841-161703}
\end{figure*}  

\begin{figure*}
    \centering
           \includegraphics[width=16cm]{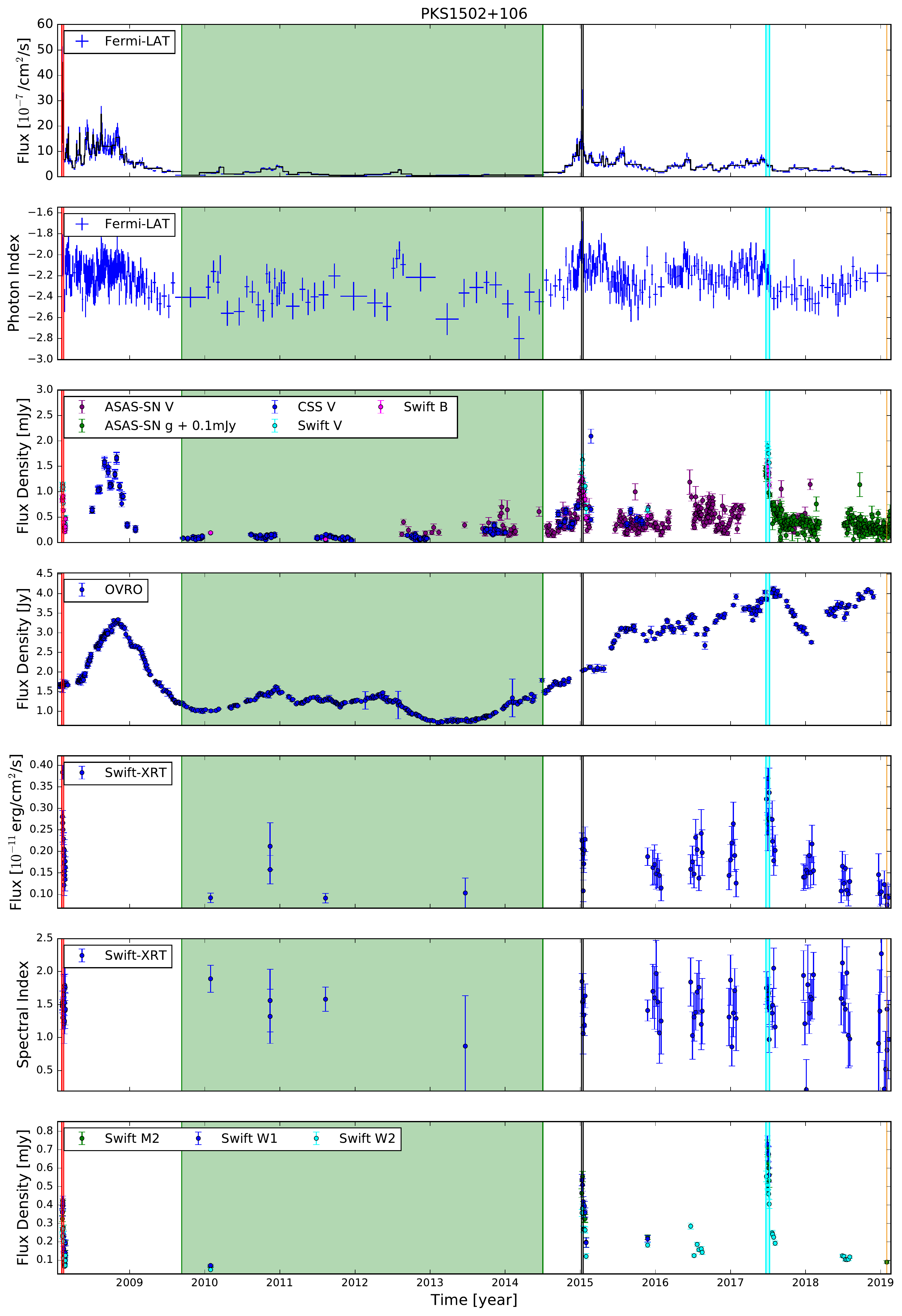}
     \caption{Multi-wavelength light curve of \PKS. The orange line indicates the arrival time of the high-energy neutrino IC-190730A. The green region in panel 1 marks the quiescent state and vertical red, cyan and black lines mark three flaring states, which are accompanied by X-ray and optical flares, selected for a dedicated gamma-ray spectral analysis (see Tab.~\ref{tab:sed} and Fig.~\ref{fig:SED}, which uses the same color code).
              }
         \label{fig:LC_PKS1502+106}
\end{figure*}

\begin{figure*}
    \centering
           \includegraphics[width=18cm]{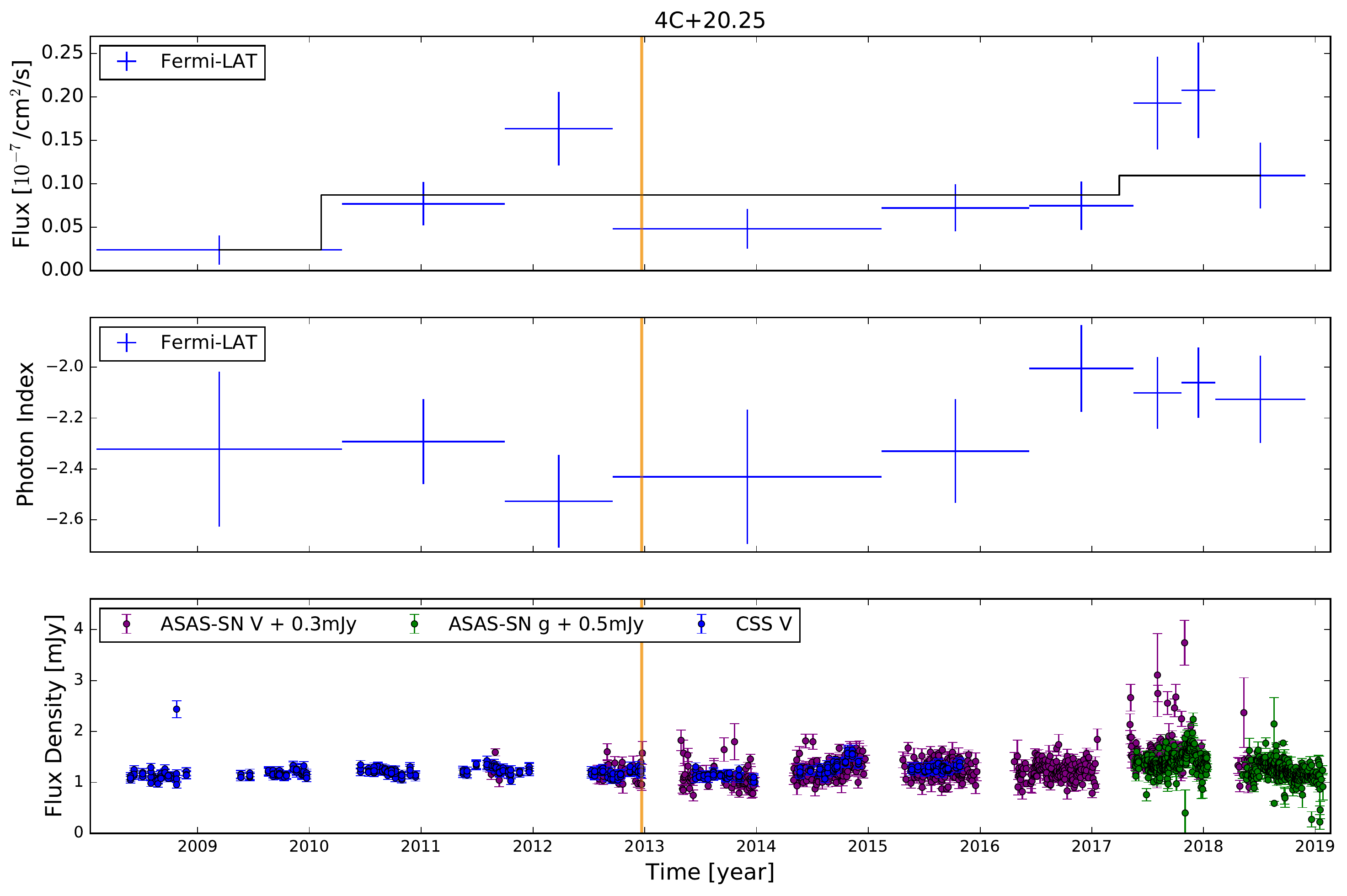}
     \caption{Multi-wavelength light curve of 4C +20.25. The duration of the neutrino flare is short ($T_w$ = 5.2\,days) and its arrival time is shown as an orange line. 
              }
         \label{fig:4C+20.25}
\end{figure*} 

\begin{figure*}
    \centering
           \includegraphics[width=18cm]{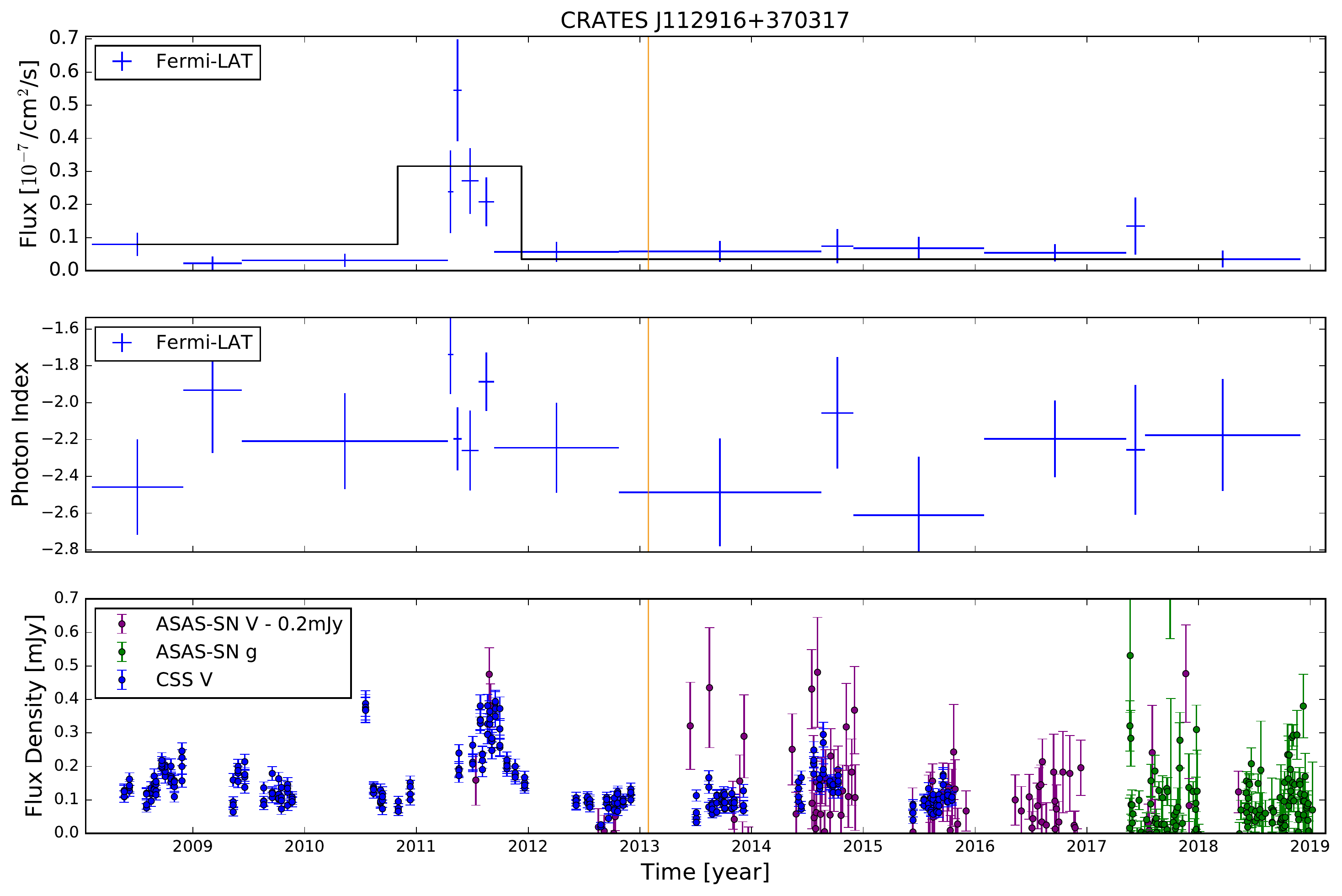}
     \caption{Multi-wavelength light curve of CRATES J112916+370317. 
     The duration of the neutrino flare is short ($T_w$ = 1.4\,h) and its arrival time is shown as an orange line. 
              }
         \label{fig:MG2J112910+3702}
\end{figure*} 

 \begin{figure*}
    \centering
           \includegraphics[width=18cm]{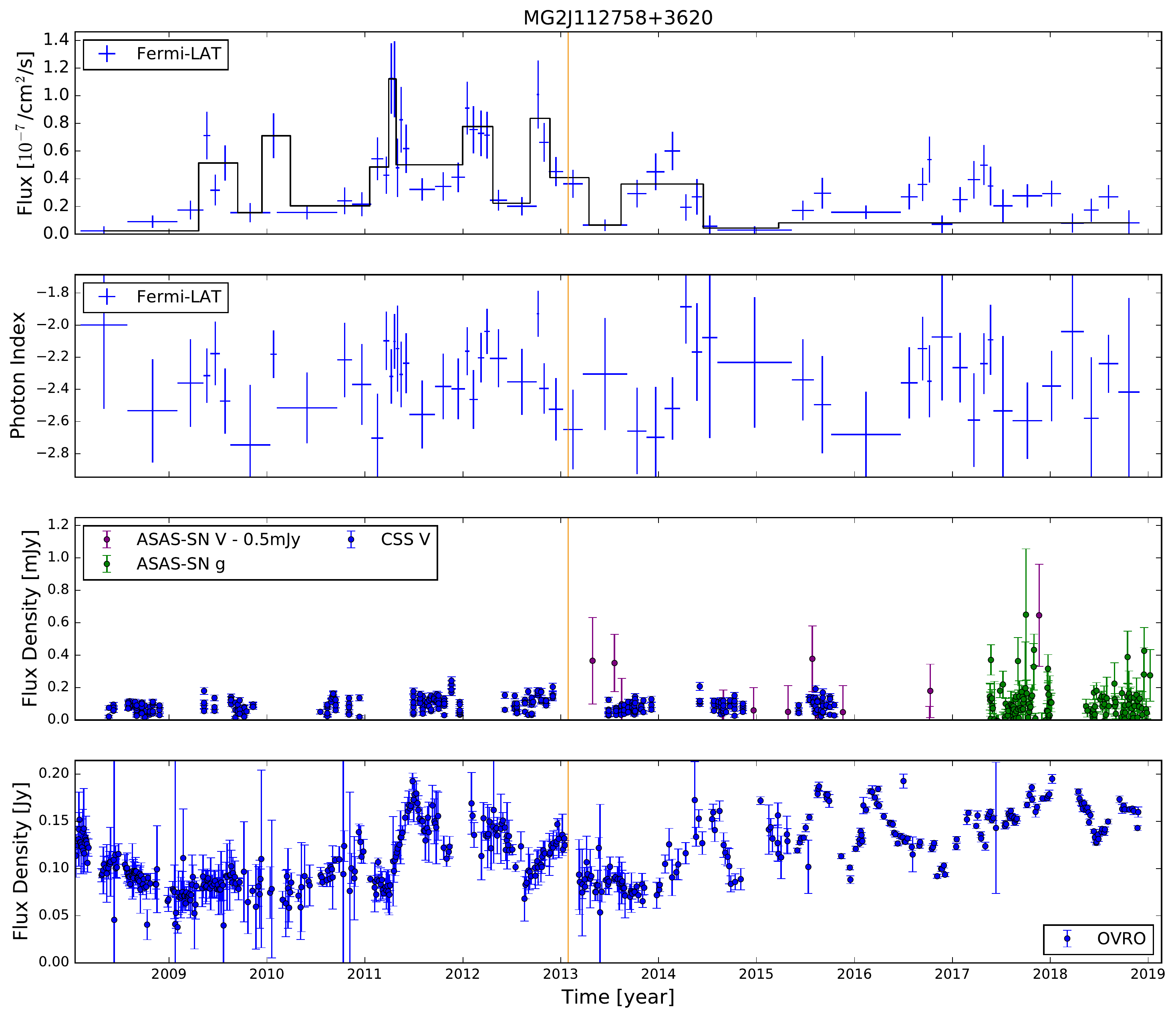}
     \caption{Multi-wavelength light curve of MG2 J112758+3620. The duration of the neutrino flare is short ($T_w$ = 1.4\,h) and its arrival time is shown as an orange line. 
              }
         \label{fig:MG2J112758+3620}
\end{figure*}

\begin{figure*}
    \centering
           \includegraphics[width=16cm]{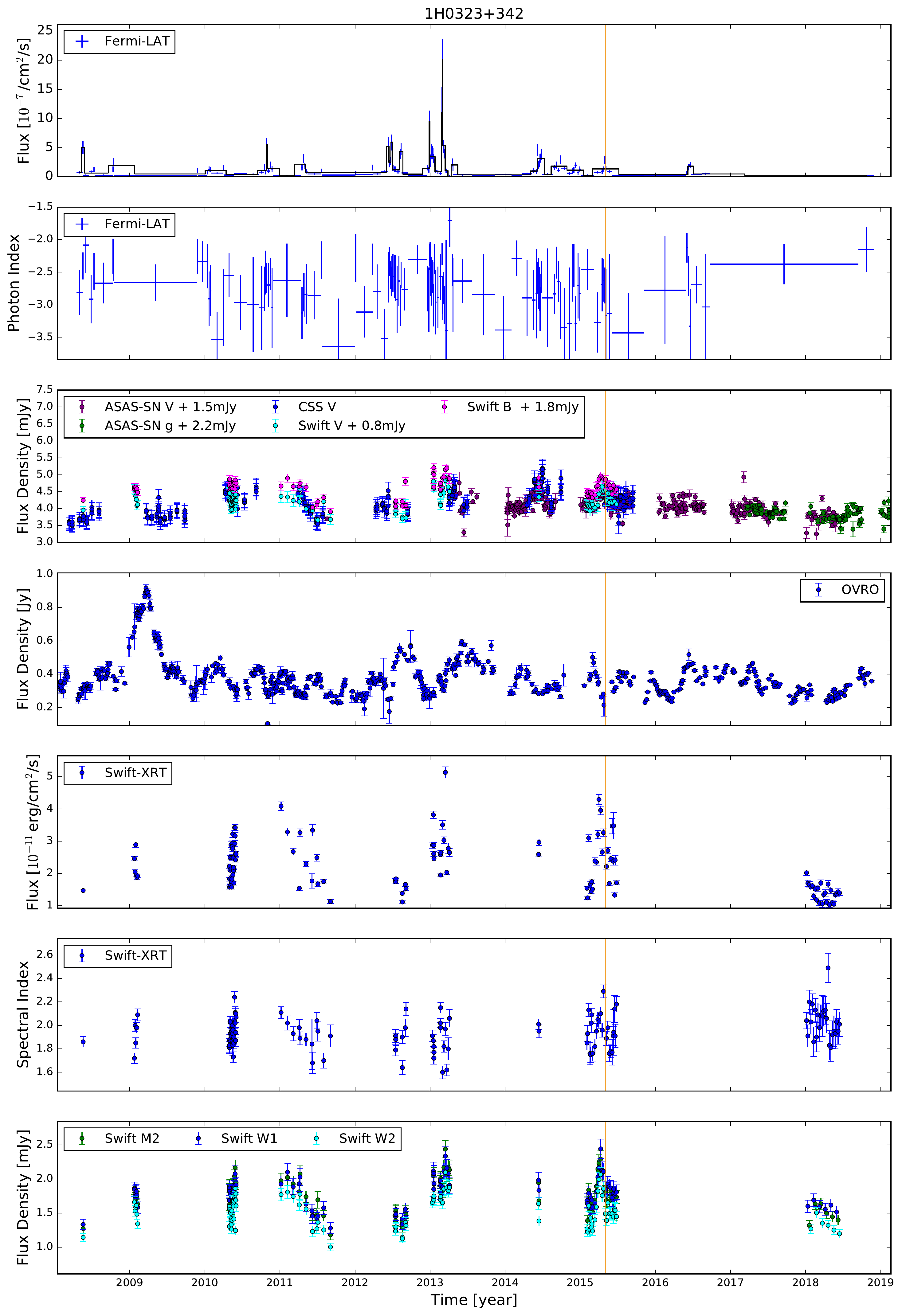}
     \caption{Multi-wavelength light curve of \HH. The duration of the neutrino flare is short ($T_w$ = 147\,s) and its arrival time is shown as an orange line.
              }
         \label{fig:LC_1H0323+342}
\end{figure*} 
 
 \begin{figure*}
    \centering
           \includegraphics[width=16cm]{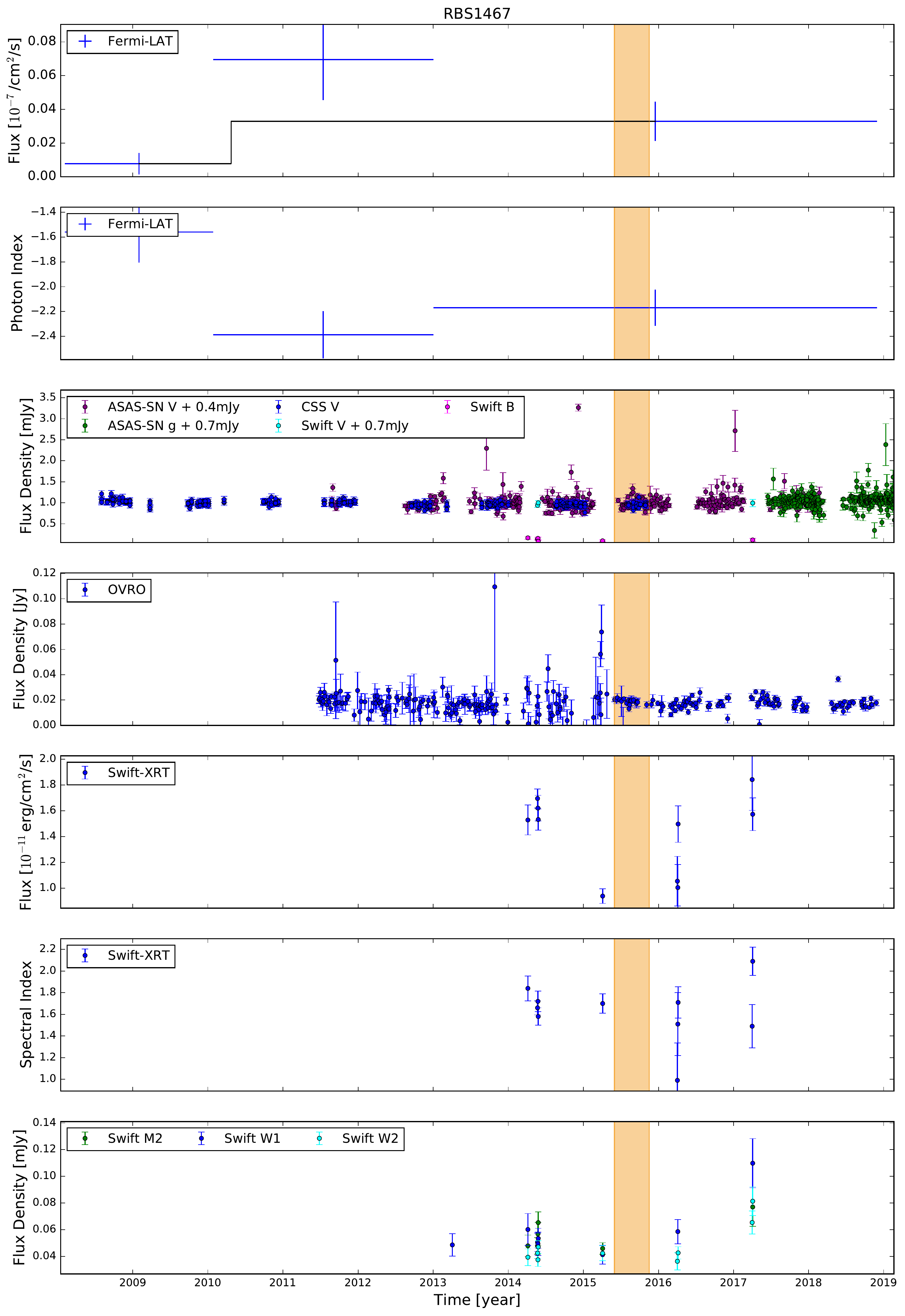}
     \caption{Multi-wavelength light curve of RBS 1467. The orange shaded region is centered on the mean of the Gaussian describing the neutrino emission with a width corresponding to twice the standard deviation.
              }
         \label{fig:RBS1467}
\end{figure*} 

 \begin{figure*}
    \centering
           \includegraphics[width=18cm]{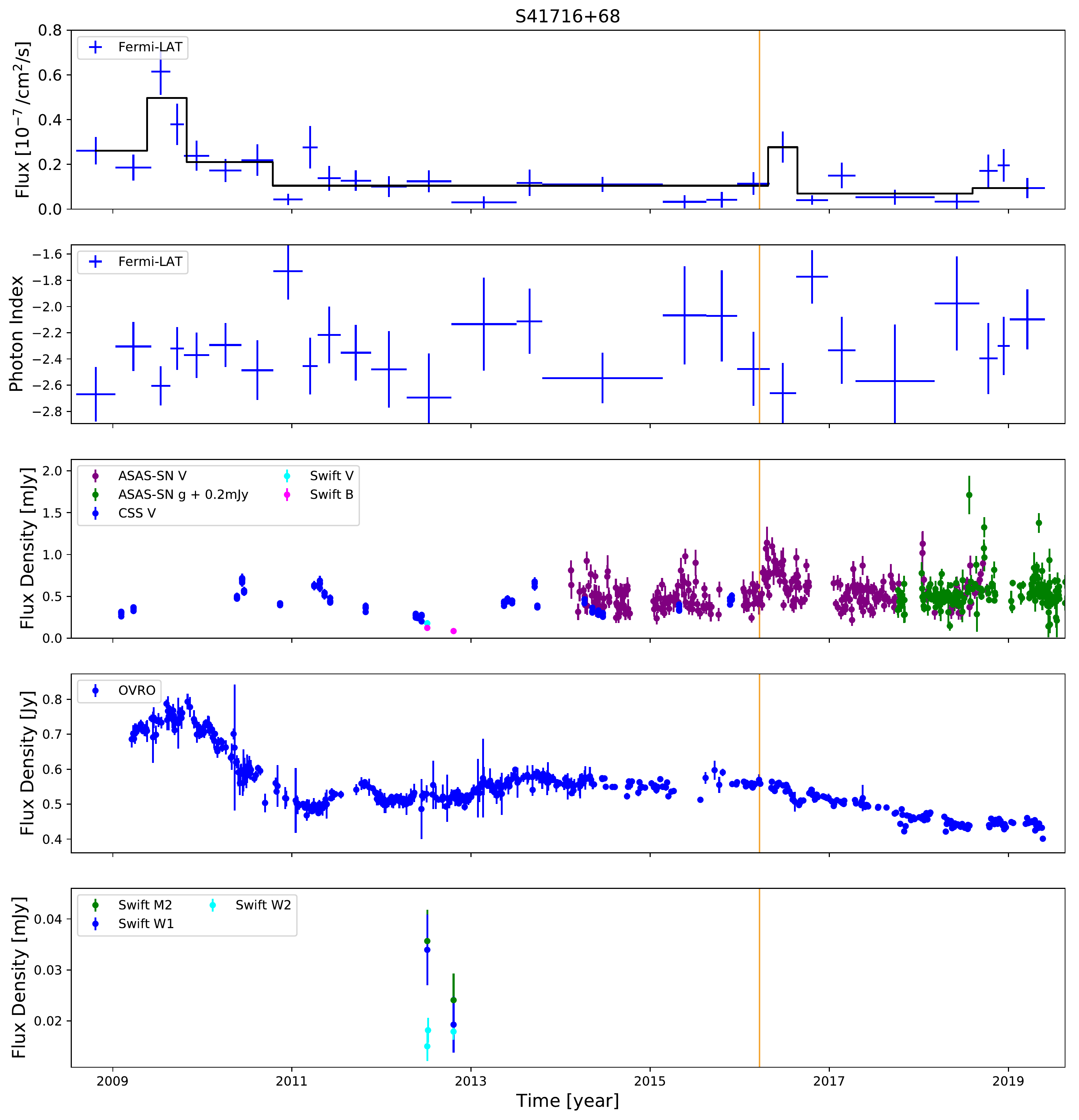}
     \caption{Multi-wavelength light curve of S4 1716+68. The duration of the neutrino flare is short ($T_w$ = 4.7\,s) and its arrival time is shown as an orange line. 
              }
         \label{fig:S41716+68}
\end{figure*} 

\begin{figure*}
    \centering
           \includegraphics[width=18cm]{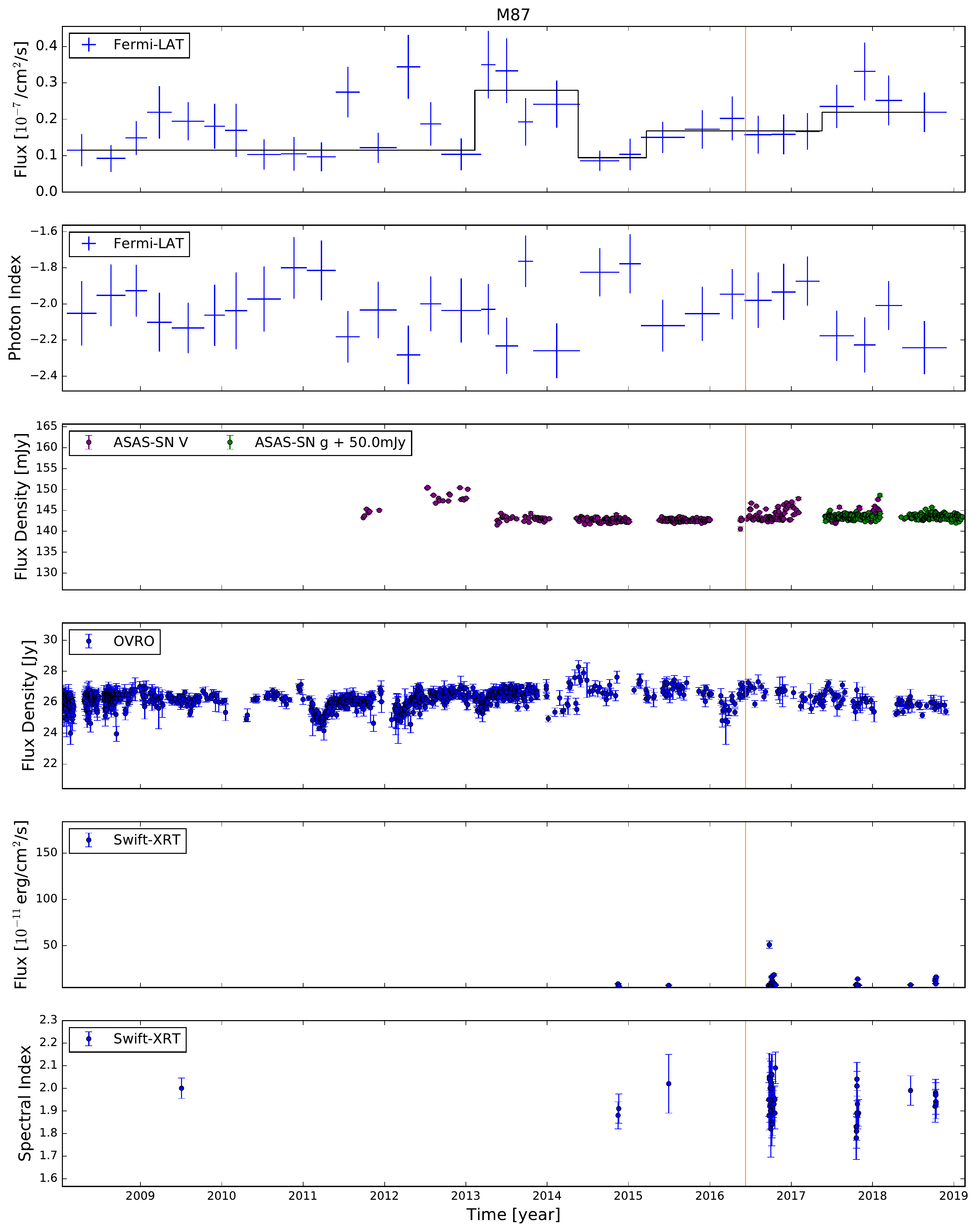}
     \caption{Multi-wavelength light curve of M87. The duration of the neutrino flare is short ($T_w$ = 3.9\,min) and its arrival time is shown as an orange line.
              }
         \label{fig:M87}
\end{figure*}

\begin{figure*}
    \centering
           \includegraphics[width=16cm]{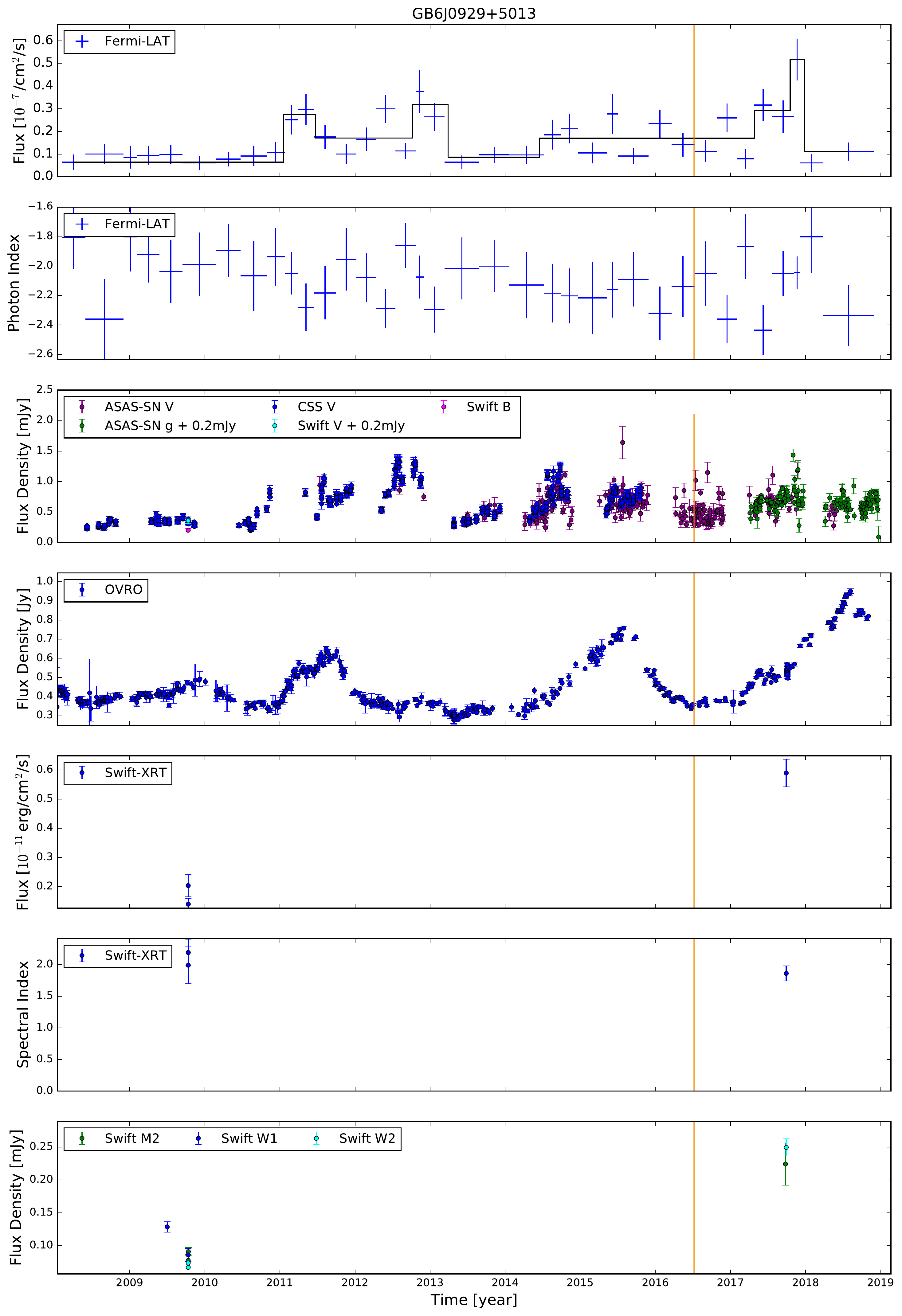}
     \caption{Multi-wavelength light curve of GB6 J0929+5013. The duration of the neutrino flare is short ($T_w$ = 1.2\,days) and its arrival time is shown as an orange line. 
              }
         \label{fig:GB6J0929+5013}
\end{figure*} 

  \begin{figure*}
    \centering
           \includegraphics[width=18cm]{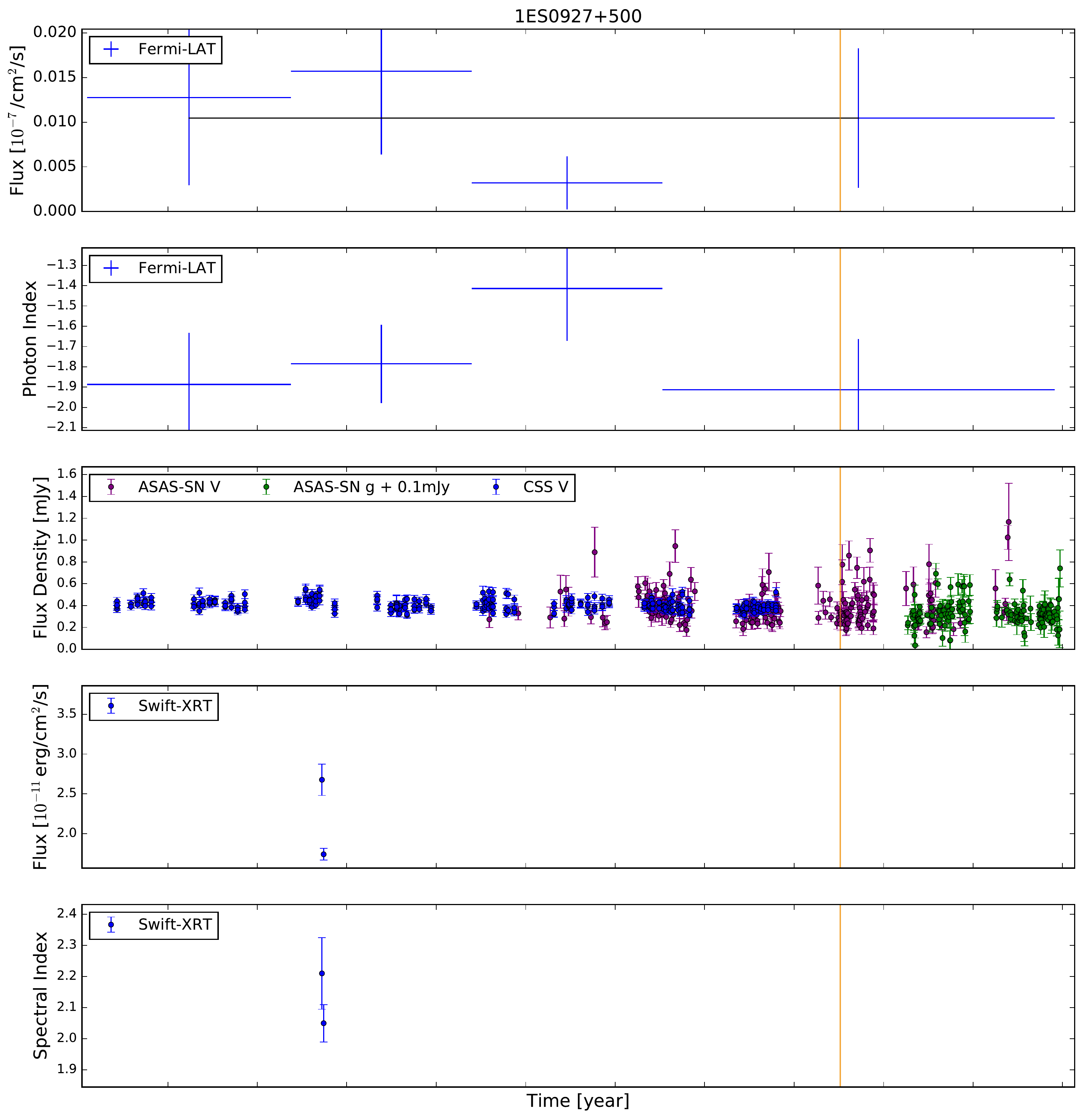}
     \caption{Multi-wavelength light curve of 1ES 0927+500. The duration of the neutrino flare is short ($T_w$ = 1.2\,days) and its arrival time is shown as an orange line. 
              }
         \label{fig:1ES0927+500}
\end{figure*}

\end{appendix}

\end{document}